\documentstyle[prb,eqsecnum,aps]{revtex}
\input epsf
\draft
\begin{document}

\title{Surface Induced Anomalous Superconductivity}

\author{ Herman J. Fink and Stephen B. Haley}

\address{Department of Electrical and Computer Engineering\\
University of California\\
Davis, CA 95616\\}

\date{\today}
\maketitle

\begin{abstract} The Ginzburg-Landau\,(GL) theory is recast using a Hamiltonian
 involving the {\it complete} kinetic energy density which requires that the 
 surface energy must contain a term $\nabla |\psi|^2$ to support superconducting
 \,(SC) states.   The GL equations contain two temperature, $t$, dependent 
 parameters $\alpha(t)$ and $\beta(t)$, which are respectively the coefficients 
 of the  SC pair density $\propto |\psi|^2$, and the pair interaction term 
 $\propto |\psi|^4$ in the free energy density.  The sign of these parameters, 
 which define distinct solution classes, and the ratio 
 $s(t) = \sqrt{|\alpha|/|\beta|}$ are governed by the characteristics of the 
 surface energy density.  In addition to the conventional bulk superconducting 
 states with ($\alpha < 0, \beta > 0)$, anomalous superconducting states exist 
 for all other sign combinations, including cases with $\beta < 0$ which may 
 exist only when surface pair interactions are significant. All possible 
 solutions of our generalized nonlinear, one dimensional GL equations are found 
 analytically and applied to a thin superconducting slab which manifests the 
 possibility of states exhibiting enhanced, diminished, and pre-wetting 
 superconductivity. Critical currents are determined as functions of $s(t)$ and 
 surface parameters. The results are applied to critical current experiments on 
 SNS systems.  \\

\end{abstract}
\pacs{    \\ } 

% body of paper
%==========================================================================
\section{INTRODUCTION}
Macroscopic effects of superconductivity are investigated using the 
Ginzburg-Landau\,(GL) phenomenological model\cite{ginzburg} for the free energy 
of the superconducting state characterized by a complex order parameter $\psi$
 and two parameters $\alpha$ and $\beta$. These parameter are, respectively, 
 coefficients of the linear superconducting\,(SC) pair density term 
 $\propto|\psi|^2$ and the quadratic pair interaction term $\propto|\psi|^4$.  
 A variational process leads to coupled GL differential equations that 
 characterize a minimum free energy state that depends on the surface energy 
 density through additional parameters.  In almost all cases the source of 
 an applied magnetic field or a current is externally controllable and the free 
 energy is of the Gibbs form.   Quantized magnetic levitation\cite{haley} is an 
 exceptional, fundamental example of a nano-magneto-mechanical device, 
 whose analysis requires the use of the Helmholtz free energy. This article 
 deals with the Gibbs free energy and the anomalous superconductivity induced 
 by externally controlled surface energy.  Since our Hamiltonian functional 
 differs from the standard GL functional, and since unexpected phenomena 
 arising from the surface energy, the article is written in a self-contained 
 format with all relations derived from our energy functional.   

In the absence of surface effects, the pair interaction parameter $\beta$ 
must be positive in order to have an energy minimum and a concomitant SC state. 
 The surface boundary condition of the superconducting order parameter modifies 
 the free energy density near the surface on the scale of the coherence length.
  For sample dimensions of $1 {\rm mm}$ and larger, boundary conditions 
  do not contribute significantly to the total free energy.  However, for a 
  SC sample, assumed to be uniform and homogeneous, with relevant dimensions of 
  $10 {\rm \mu m}$ and smaller, the total free energy can be significantly 
  altered by boundary effects. 

For a weakly superconducting surface, considered previously
 \cite{degennes,simonin,indekeu}, the surface energy term added to the 
 Ginzburg-Landau (GL) free energy, has the form 
 $\propto\int_s(|\psi|^2/b_2)ds $, where the integral is extended over 
 the sample surface, and $b_2$ is a characteristic length.  When $b_2 < 0$, 
 the overall free energy is lowered and an increase of the transition 
 temperature above the bulk value $T_c$ is possible\cite{fink}.  If $b_2 > 0$, 
 the free energy is increased and the nucleation temperature is lowered 
 relative to $T_c$.  In the weak SC surface limit, the sample is 
 superconducting only if $\beta$ is positive. 

If the surface is strongly superconducting, the surface energy has an 
additional SC pair interaction term $\propto\int_s(|\psi|^4/b_4)ds$, with $b_4$ 
a characteristic length.  For $b_4 > 0$, we show that the entire sample may be 
in an anomalous superconducting state with $\beta$ negative.  We also show that 
the minimum free energy explicitly depends on $b_4$; whereas the dependence on 
$b_2$ is only implicit through $|\psi|$.  Thus, the parameter $b_4$ essentially 
determines whether a particular solution of the differential GL equations 
represents a physical SC state, or whether it is merely a mathematical solution.

To account for the spatial variation of $\psi$, the conventional GL energy 
functional includes the term $\propto |{\bf p}\psi|^2$, where ${\bf p}$ is 
the momentum operator. In this study we introduce a Hamiltonian with the 
complete kinetic energy density 
$K \propto \psi^*{\bf p}^2\psi = 
|{\bf p}\psi|^2-i\epsilon\nabla\cdot(\psi^*{\bf p}\psi)$, 
where the parameter $\epsilon = 1$, or $0$ is used to keep track of the 
difference between the $K$ model and the standard GL model.  In view of the 
divergence operation, the additional term is a surface term. The imaginary 
part of the divergence term is identically zero, but the real part changes 
the surface boundary conditions, and hence the character of the order parameter
 throughout the sample volume. It is shown that the real part of the divergence 
 term in the $K$ model destroys superconductivity unless the surface energy is 
 augmented by a term $c|\psi|\nabla|\psi|$, where $c \ge \epsilon$ is a 
 constant. Furthermore, a second order phase transition is possible only if 
 $c = \epsilon$, and the transition temperature is higher for $c = \epsilon$ 
 than for any value of $c > \epsilon$.  For $c < \epsilon$ SC states do not 
 exist for small values of the order parameter;  however, first order phase 
 transitions may exist.      
 
In the GL theory $\alpha$ and $\beta$ are arbitrary parameters, assumed to be 
functions of the reduced temperature $t = T/T_c$. However, the explicit 
temperature dependence of $\alpha$ and $\beta$ is not given and thus has 
to be conjectured from experiments.  Since we do not want to assume 
{\it a priori} an explicit temperature dependence, we first investigate the 
solutions of the GL equations consistent with the boundary conditions, 
independent of specific temperature models for $\alpha$ and $\beta$.   
For one dimensional systems with a uniform current density j, all possible 
physical solutions of the GL equations, with sgn$(\alpha ) = \pm 1$ and 
sgn$(\beta) = \pm 1$ are found, and categorized. These solutions are then 
applied to a plane slab SC sample and the influence of the surface energy 
parameters $b_2$ and $b_4$ on the sample order parameter is studied, and 
unusual SC states, including ``pre-wetting'' surface states, are analyzed. 
Critical currents are shown as functions of $s(t)$.  A temperature model is 
introduced for the fundamental ratio $s(t) = \sqrt{|\alpha(t)|/|\beta(t)|}$ 
to illustrate possible enhancement or reduction of the critical temperature. 
 Finally, the critical current of a superconductor-normal-superconductor (SNS) 
 layered system is calculated, and the results with $\alpha > 0, \beta > 0$ 
 are shown to be in excellent agreement with critical current 
 experiments.\cite{dubos}

%===========================================================================
\section{GENERAL EQUATIONS}

As originally formulated by Ginzburg and Landau\cite{ginzburg}, the free energy 
of a superconductor in a magnetic field ${\bf H}$ is a function of a coordinate 
dependent complex order parameter $\psi ({\bf r})$ and the vector potential 
${\bf A}({\bf r})$. We begin our development with a general energy functional 
$G$ that encompasses a broad spectrum of models, including those leading to the 
stationary states of the Schr\"{o}dinger equation and the GL equations for SC. 
It is

\begin{equation}\label{G}
G = \int_V d^3{\bf r}\left[U({\bf r}, |\psi({\bf r})|)  + K({\bf r}) + 
{1\over 2\mu_o}[\nabla\times{\bf A}'({\bf r})]^2\right] + 
{\hbar^2\over 2m^*}\int_S d{\bf s}\cdot[\hat{\bf n}\Lambda(\psi,{\bf A})+ 
{c\over 2}\nabla|\psi|^2], 
\end{equation}
where $U$ is a potential energy density that depends only on functions of 
${\bf r}$ and $|\psi({\bf r})|$, $K({\bf r})$ is the kinetic energy density, 
and ${\bf A}'$ is a linear function of ${\bf A}$.  Specific forms of $U$ and 
${\bf A}'$ will be detailed later.  The second integral is over the surface 
$S$ enclosing the volume $V$. The unit vector $\hat{\bf n}$ is normal to the 
surface at each point on S.  The surface integrand function $\Lambda$, which 
has units of (length)$^{-4}$, is assumed to contain all parameters that 
characterize not only the intrinsic aspects of the surface, but also external 
effects that link to the sample volume via the surface.  It is shown below that 
the $c\nabla|\psi|^2$ term is necessary for a SC state to exist, and the value 
of c determines whether a SC state exists and whether a phase transition to the 
SC state is first or second order.  The boundary conditions for $\psi$ depend 
on $c$ and $\Lambda$, and a specific form for $\Lambda$ is introduced. 

The kinetic energy density $K$ of a particle of effective mass $m^*$ and charge 
magnitude $e^*$ subject to a vector potential ${\bf A}$ is given by
\begin{equation}\label{K}
{2m^*\over \hbar^2}K({\bf r}) = \psi^*{\bf p}^2\psi = 
|{\bf p}\psi|^2-i\epsilon\nabla\cdot(\psi^*{\bf p}\psi),
\end{equation}
where
\[{\bf p} = -i\nabla + {e^*\over \hbar}{\bf A},\]
The first term in Eq.\ (\ref{K}) is the form used in the standard GL functional.
 To distinquish the effects arising from the complex divergence term the 
 parameter $\epsilon = 1$ or $0$ is introduced. When $\epsilon = 0$ one 
 recovers the standard GL model. As shown below, only the real part of the 
 divergence term in $K$ contributes to $G$.  
 
Introducing the modulus-phase form
\begin{equation}\label{modphase}
 \psi({\bf r}) = |\psi({\bf r})|\exp [i\theta ({\bf r})],
\end{equation}
the density $K$ assumes the form
\begin{equation}\label{KQ}
{2m^*\over \hbar^2}K({\bf r}) = [\nabla|\psi|\cdot\nabla|\psi| + 
|\psi|^2{\bf Q}\cdot{\bf Q}] - i\epsilon\nabla\cdot[|\psi|^2{\bf Q} - 
{i\over 2}\nabla|\psi|^2],
\end{equation} 
where the vector ${\bf Q}$ is the gauge invariant phase gradient
\begin{equation}\label{Q}
{\bf Q} = \nabla\theta + {2\pi\over\phi_o}{\bf A},
\end{equation}
with $\phi_o = h/e^*$ the flux quantum. In the superconducting case,  
$e^* = 2|e| > 0$ is the SC pair charge.  The magnetic field ${\bf H}$ is 
given by
\begin{equation}\label{H}
\mu_o{\bf H} = \nabla\times{\bf A} = 
{\phi_o\over 2\pi}\nabla\times({\bf Q}-\nabla\theta).
\end{equation}
As will be shown, the term $\nabla\cdot[|\psi|^2{\bf Q}] = 0$ 
in Eq.\ (\ref{KQ}), since it is proportional to the divergence of the real 
current density.  The term $\nabla\cdot\nabla|\psi|^2$ may appear to be the 
divergence of a diffusion current, but it is not.  It contributes to the 
surface energy.   Using Eq.\ (\ref{KQ}), the functional $G$ is
\begin{eqnarray}\label{G1}
G & = &\int_V d^3{\bf r}\left[U(|\psi|)  + 
{\hbar^2\over 2m^*}(\nabla|\psi|\cdot\nabla|\psi| + 
|\psi|^2{\bf Q}\cdot{\bf Q}) + 
{1\over 2\mu_o}[\nabla\times{\bf A}']^2\right] + \\
& &{\hbar^2\over 2m^*}\int_S d{\bf s}\cdot[\hat{\bf n}\Lambda(|\psi|, \theta, 
{\bf A}) + {1\over 2}(a - 1)\nabla|\psi|^2],~~~~\mbox{with}~~~a = 
1 + c - \epsilon. \nonumber 
\end{eqnarray}
 
The conventional GL free energy difference $\Delta G = G_{sc} - G_{n}$ 
between the superconducting and normal states is given by $G$ in Eq. (\ref{G1}) 
with $c = \epsilon = 0$ and  ${\bf A}' = {\bf A}-{\bf A}_a$, 
where ${\bf A}_a$ is the vector potential of the applied field ${\bf H}_a$. 
 Note that for the standard GL model $\epsilon = 0$ in 
$K \propto \psi^*{\bf p}^2\psi = 
|{\bf p}\psi|^2-i\epsilon\nabla\cdot(\psi^*{\bf p}\psi)$.
The vector potential ${\bf A}$ that appears in ${\bf Q}$ reduces to ${\bf A}_a$ 
for single particle models, since a single charged particle does not interact 
with its own field.

The vectors $\nabla\theta$ and ${\bf A}$ in ${\bf Q}$ are distinct mathematical 
entities.  When a magnetic field is present $\nabla\times{\bf A}\ne 0$, whereas 
$\nabla\times\nabla\theta = 0$, except at vortex centers, where 
$\nabla\times\nabla\theta$ is a sum of delta functions\cite{fath,haley1}.  
Thus $\theta$ and ${\bf A}$ must be considered as independent variables, 
and the minimum set of independent variables contains three elements, 
which we chose as $(|\psi|, \theta, {\bf A})$.   As outlined the Appendix, 
setting the variation $\delta G = 0$, leads to the differential equations in 
the volume and corresponding surface boundary conditions.  The variation with 
respect to $|\psi|, \theta$ and ${\bf A}$, respectively yield 

\begin{equation}\label{geq1}
\nabla^2 |\psi| = {m^*\over \hbar^2}\frac{\partial U}{\partial |\psi|} +
 |\psi| {\bf Q}\cdot{\bf Q},
\end{equation}

\begin{equation}\label{geq2}
\nabla\cdot(|\psi|^2{\bf Q}) = 0,
 \end{equation}

\begin{equation}\label{geq3}
{1\over \mu_o}\nabla\times(\nabla\times{\bf A}')  =  
-\epsilon_s{e^*\hbar\over m^*}|\psi|^2{\bf Q}.
\end{equation}
The parameter $\epsilon_s = 0$ for single particle models and 
$\epsilon_s = 1$ for models that include particle interaction terms, 
e.g. the GL model. 

The corresponding surface boundary conditions are 
\begin{equation}\label{gbc1}
[(a + 1)\nabla |\psi|+(a - 1)|\psi|\frac{\partial\nabla |\psi|}{\partial |\psi|}
]\cdot\hat{\bf n} = -{\partial \Lambda\over\partial |\psi|} ,
\end{equation}
\begin{equation}\label{gbc2}
2|\psi|^2{\bf Q}\cdot\hat{\bf n}  = {\partial \Lambda\over\partial\theta},
\end{equation}
and
\begin{equation}\label{gbc3}
{1\over \mu_o}\hat{\bf n}\times(\nabla\times{\bf A}') = 
{\hbar^2\over 2m^*}\nabla_{\bf A}\Lambda.
\end{equation}
As outlined in the last paragraph of the Appendix, the minimum energy 
functional consistent with Eq.\ (\ref{geq1})- (\ref{gbc3}) assumes the form

\begin{equation}\label{Gmin}
G_{min} = \int_V d^3{\bf r} \left\{U- {1\over 2}|\psi|\frac{\partial U}
{\partial |\psi|} + {1\over 2\mu_o}[\nabla\times{\bf A}']^2\right\}  + 
{\hbar^2\over 2m^*}\int d{\bf s}\cdot[\hat{\bf n}\Lambda + 
{1\over2}a\nabla|\psi|^2].
\end{equation}
Using the boundary condition (\ref{gbc1}) with $a = 1$, the expression for 
$G_{min}$ can be written in the form
\begin{equation}\label{Gmin1}
G_{min} = \int_V d^3{\bf r} \left\{g(U) + {1\over 2\mu_o}
[\nabla\times{\bf A}']^2\right\}  + 
{\hbar^2\over 2m^*}\int d{\bf s}\cdot\hat{\bf n}g(\Lambda),
\end{equation}
where
\[g(F) = -{1\over 2}|\psi|^3\frac{\partial}{\partial |\psi|}\left[
{F\over |\psi|^2}\right] .\]
Equation (\ref{Gmin1}) shows the functional symmetry of the volume potential 
energy density $U$ and the surface potential energy per unit area 
$(\hbar^2/2m^*)\Lambda$ which determine $G_{min}$.  It is evident from this 
form of $G_{min}$ that if $U$ and $\Lambda$ are expanded in powers of the 
particle density, the coefficient of leading term $|\psi|^2$ does not contribute 
directly to $G_{min}$.  To proceed further one must introduce models  for the 
functions $U$ and $\Lambda$.

For the potential energy density $U$ we assume the form

\begin{equation}\label{Upot}
U({\bf r}, |\psi({\bf r})|) = \alpha({\bf r}) |\psi({\bf r})|^2 + 
{1\over 2}{\beta({\bf r})\over n_s^*}|\psi({\bf r})|^4,
\end{equation}
where $n_s^*$ is a constant reference particle density.  As formulated, 
the parameters $\alpha, \beta,$ have units of energy. The function $U$ in  
(\ref{Upot}) encompasses both the Schr\"{o}dinger quantum stationary state 
and GL superconductivity models.  In the former case, $\alpha({\bf r}) = 
V({\bf r})- E$, and $\beta = 0$, with $E$ the energy eigenvalue and $V$ the 
potential.  For the GL model,  $\alpha$ and $ \beta$ are independent of 
${\bf r}$, $n_s^*$ is the SC pair density at $T = 0$, and  $\psi({\bf r})$ 
is the complex order parameter.

Since it is generally easier to solve the linear Schr\"{o}dinger equation for 
the complex wave function $\psi$ than the coupled nonlinear Eqs.\ (\ref{geq1}) 
and (\ref{geq2}), we do not pursue the coupled approach here;  however, there 
are some advantages to this approach for interacting electron 
systems\cite{fath,haley1}.  The remainder of the paper concerns superconducting 
systems with $\alpha$ and $ \beta$ independent of the coordinate ${\bf r}$.

For mathematical expediency it is convenient to  normalize the order parameter 
using a positive constant $\psi_o$.  Since the GL energy and resulting GL 
variational equations are independent of the value of $\psi_o$, its choice is 
arbitrary.   Using the scaled temperature notation $t = T/T_c$, where 
$T_c$ is the temperature at which a second order phase transition, with limit 
$|\psi| \rightarrow 0$, would occur in a bulk sample, we employ the implicit 
temperature dependent normalization
\begin{equation}\label{fdef}
f({\bf r}) = {|\psi({\bf r})|\over \psi_o(t)},~~~~\psi_o(t) = 
\sqrt{n_s^*}s(t).
\end{equation} 
The unitless temperature dependent function $s(t)$ is defined by the 
fundamental length ratio
\begin{equation}\label{s}
s(t) = {l_{\beta}(t)\over l_{\alpha}(t)} = \sqrt{{|\alpha(t)|\over|\beta(t)|}},
\end{equation}
where the lengths are defined as 
\begin{equation}\label{lengths}
l_{\alpha} = {\hbar\over\sqrt{2m^*|\alpha|}},~~~~l_{\beta} = 
{\hbar\over\sqrt{2m^*|\beta|}}.
\end{equation}
In conventional superconductivity, with $(\alpha < 0, \beta > 0)$, the length 
$l_{\alpha}$ is referred to as the GL coherence length, usually denoted as 
$\xi$. For a bulk superconductor with $\Lambda = 0$
the pair density for minimum energy $|\psi_{m}(t=0)|^2 = 
-{{\alpha(0)/\beta(0)}} = 1$. Thus the maximum value of 
$s(t)$ is $s(0) = 1$, i.e. $ l_{\alpha} (0) = l_{\beta} (0)$.  When 
$\Lambda \neq 0$, anomalous SC states exist with $s(t) > 1$, even when $t > 0$; 
thus we retain the notation $l_{\alpha}$ to denote a fundamental length that 
can assume values outside the range of the conventional coherence length $\xi$.
We also define a normalized, unitless, order parameter
\begin{equation}\label{chi}
\chi ({\bf r}) = {|\psi({\bf r})|\over \sqrt{ n_s^*}} = s(t)f({\bf r}).
\end{equation}
Because the scaling parameter $n_s^*$ is temperature independent, 
$\chi$ is always physically meaningful; whereas $f$ may become singular when 
the temperature dependent scaling parameter $s(t) \rightarrow 0$. Thus the 
function $f$, although mathematically useful, is not in general a physical 
order parameter.

Since $\Delta G$ depends not only on the magnitudes, but also the signs of 
$\alpha$ and $\beta$, it is convenient to write $\alpha = 
|\alpha |\mbox{sgn}(\alpha)$, and $\beta = |\beta|\mbox{sgn}(\beta)$, 
where sgn$(...) = \pm 1$. Using Eqs.\ (\ref{Upot})- (\ref{lengths}), the 
equations (\ref{geq1})-(\ref{geq3}) in the volume V assume the form

\begin{equation}\label{gl1n}
l_{\alpha}^2\nabla^2 f = [\mbox{sgn}(\alpha) + \mbox{sgn}(\beta)f^2 + 
l_{\alpha}^2 {\bf Q}\cdot{\bf Q}]f,
\end{equation}

\begin{equation}\label{divj}
\nabla\cdot(f^2{\bf Q}) = 0,
\end{equation}

\begin{equation}\label{gl2n}
\nabla\times [\nabla\times({\bf Q}-{\bf Q}_a)] + 
\left({s\over \lambda_L}\right)^2 f^2 {\bf Q} = 
\nabla\times[\nabla\times\{\nabla (\theta -\theta_a)\}].
\end{equation}
with $\lambda_L = (1/e^*)\sqrt{m^*/(\mu_o n_s^*)}$ the London magnetic field 
penetration depth which is temperature independent.
The inhomogeneous term in Eq.\ (\ref{gl2n}) is present when 
vortices, or 3D worm holes exist\cite{haley1}.  Equation (\ref{gl2n}) may be
 written in the form
\begin{equation}\label{gl2}
\nabla\times({\bf H}-{\bf H}_a) = 
- {\hbar\over e^*\mu_o}\left({s\over \lambda_L}\right)^2 f^2 {\bf Q} = 
-{e^*\hbar\over m^*} |\psi|^2{\bf Q}.
\end{equation}
The right hand side of Eq.\ (\ref{gl2}) is the physical super current density 
${\bf j}$, which we write in terms of a unitless current density parameter 
${\bf J}$ by
\begin{equation}\label{j}
{\bf j} = -j_o{\bf J}, ~~~~\mbox{where}~~~~~~{\bf J} = 
l_{\alpha} f^2{\bf Q},~~~~~~j_o = 
{\hbar\over e^*\mu_ol_{\alpha}}\left({s\over \lambda_L}\right)^2 .
\end{equation}
Although ${\bf J}$ is a convenient parameter for mathematical reasons, 
it is not appropriate for physical interpretation since it is singular in the 
limit $s \rightarrow 0$.  Thus we also define the unitless current parameter
\begin{equation}\label{kj}
{\bf k}_j = {m^*l_{\beta}\over e^*\hbar n_s^*}{\bf j} = -s^3(t){\bf J}.
\end{equation}  
For plotting, ${\bf k}_j$ is a well behaved parameter that tracks the physical 
current density ${\bf j}$.

The surface boundary conditions from Eqs.\ (\ref{gbc1}-)(\ref{gbc3}) are  

\begin{equation}\label{bc1}
[(a+1)\nabla f + (a-1)f\frac{\partial\nabla f}{\partial f}]\cdot\hat{\bf n} = 
-{1\over \psi_o^2}{\partial \Lambda(f, \theta)\over\partial f},
\end{equation}
where $a = 1+c-\epsilon$, and
\begin{equation}\label{bc2}
\hat{\bf n}\times({\bf H}-{\bf H}_a) = 
{\hbar^2\over 2m^*} \nabla_{\bf A} \Lambda,
\end{equation}

\begin{equation}\label{bc3}
{2m^*\over e^*\hbar}{\bf j}\cdot\hat{\bf n} = 
-2\psi_o^2 f^2{\bf Q}\cdot\hat{\bf n}  = 
{\partial \Lambda(f, \theta)\over\partial\theta}.
\end{equation}
Equation (\ref{divj}) is the continuity equation for ${\bf j}$, and 
Eq.\ (\ref{bc3}) is the boundary condition for the current normal to the 
surfaces. If $\Lambda(f, \theta)$ contains a product of $f$ and $\theta$, 
then the boundary conditions (\ref{bc1}) and (\ref{bc3}) are coupled.

From Eq.\ (\ref{Gmin}), the minimum free energy $\Delta G = G_{sc} - G_{n}$ for 
superconducting systems assumes the form

\begin{equation}\label{delGmin}
\Delta G_{min} =  {1\over 2} \int_V d^3{\bf r} 
\left[({\bf A}-{\bf A}_a)\cdot{\bf j}  - 
\mbox{sgn}(\beta) n_s^*|\beta|s^4f^4({\bf r}) \right]  + 
{\hbar^2\over 2m^*}\int d{\bf s}\cdot[\hat{\bf n}\Lambda + 
{1\over2}a n_s^*s^2\nabla f^2].
\end{equation}

To proceed further one must formulate a model for the function 
$\Lambda(f, \theta)$, which determines the boundary conditions for $\psi$ 
and the minimum free energy. When currents are present a general form for 
$\Lambda(f, \theta)$ is not known.  However, if the surface is that of a SNS 
junction, or Josephson junction, with tunneling supercurrent density 
${\bf j}\cdot\hat{\bf n} = j_c\sin\Delta\theta$, it follows from 
Eq.\ (\ref{bc3}) that $\Lambda$ must contain a term 
$\propto -j_c\cos\Delta\theta$, where $\Delta\theta$ is the phase difference 
across the link.
  
In the absence of transport currents, the superconducting sample in volume $V$ 
is in thermal equilibrium with the surrounding material.  One may assume that 
the sample and its surface are characterized by an order parameter $\psi$, with 
$\Lambda$ a function of $|\psi| = \psi_o f$ only.   As discussed after 
Eq.\ (\ref{Gmin1}), the surface energy term $\Lambda$ contributes to 
$\Delta G_{min}$ only if $\Lambda $ deviates from the form 
$\Lambda \propto |\psi|^2$.  Thus, to obtain the leading surface contribution 
for ${\bf J} = 0$, we introduce the surface energy function $\Lambda$ as
\begin{equation}\label{Lambdas}
\Lambda = {1\over b_2}|\psi|^2 + {1\over 2b_4n_s^*}|\psi|^4 = 
\psi_o^2\left({1\over b_2}f^2 + {1\over 2b_4}s^2f^4\right).
\end{equation}
The first term is the form introduced by de Gennes\cite{degennes}, where 
$b_2$ is a characteristic length for the surface. The second term represents 
the superconducting pair interaction energy, with $b_4$ a characteristic 
length.  This term must be present in our approach for temperatures 
well below the critical temperature to be consistent with the 
retention of the $f^4$ term in the bulk free energy. 
The $f^4$ term may also be significant over most of the 
temperature range when the surface is strongly superconducting due to contact 
with a higher $T_c$ superconductor than the sample considered.  Substituting 
the surface model (\ref{Lambdas}) into Eq.\ (\ref{delGmin}), and using the 
boundary condition (\ref{bc1}) to eliminate $1/b_2$ leads to the form 
\begin{equation}\label{GminMod}
\Delta G_{min} =  - {1\over 2} n_s^*|\beta|s^2\left\{\mbox{sgn}(\beta)s^2 
\int_V d^3{\bf r}f^4({\bf r}) + 
l_{\beta}^2\int d{\bf s}\cdot f^2\left[(a-1)\left(\frac{\partial\nabla f}
{\partial f} - \frac{\nabla f}{f}\right) + 
\hat{\bf n}{1\over b_4}s^2f^2 \right]\right\}.
\end{equation} 
Although the volume term with $\alpha$ cancelled and the surface term with 
$b_2$ was eliminated, $\Delta G_{min}$ remains an implicit function of 
$\alpha$ and $b_2$ since $f$ depends on them via Eqs.\ (\ref{gl1n}) and 
(\ref{bc1}).  If $a \ne 1 ~ (c \ne \epsilon)$, it will be shown that 
$\Delta G_{min}$ is greater than it is for $a = 1$.  Furthermore, only $a = 1$ 
permits a second order phase transition.  For $a = 1$ and  
$l_{\beta}^2/b_4 \ge 1$ the parameter $\beta$ can be negative and still have 
$\Delta G_{min} < 0$, i.e. the overall sample is in SC state.  
If $l_{\beta}^2/b_4 \ll 1$, the surface effect is negligible and the parameter 
$\beta$ must be positive for a SC state to exist.  In the Gor'kov 
derivation \cite{gorkov} of the GL equations from the microscopic BCS 
theory\cite{BCS}, $\beta$ is always positive.  However, surface effects were 
not considered in the Gor'kov derivation.

As final observations for Section II, we examine the scaling roles of the 
fundamental lengths $l_{\alpha}$ and $l_{\beta}$, and the uniform solution of 
Eq.\ (\ref{gl1n}).  It is evident from Eqs.\ (\ref{gl1n}), (\ref{gl2n}), and 
the definition of ${\bf Q}$ that $l_{\alpha}$ scales the coordinate ${\bf r}$, 
i.e. one may use ${\bf r} \rightarrow {\bf r}/l_{\alpha}$.  Multiplying 
Eq.\ (\ref{gl2n}) by $l_{\alpha}^3$ to scale the coordinate, the only parameter 
in Eq.\ (\ref{gl2n}) is the experimentally observable parameter $\kappa = 
\lambda_s /l_{\alpha} = \lambda_L /l_{\beta}$, where  $\lambda_s = 
\lambda_L /s$ is equal to the Ginsburg-Landau penetration depth $\lambda_{GL}$ 
in conventional superconductivity.  However, for the anomalous SC cases 
analyzed in section III, $\lambda_s$ can lie outside the range of the 
conventional $\lambda_{GL}$.  The lengths $l_{\alpha}$ and $l_{\beta}$ defined 
in Eq.\ (\ref{lengths}) are positive real numbers.  All signs consistent with 
the general GL Eq. (\ref{gl1n}) are in $\mbox{sgn}(\alpha) = \pm 1$, and 
$\mbox{sgn}(\beta) = \pm 1$, which dictate whether a solution $f$ is an 
oscillatory or non-oscillatory function of the coordinate ${\bf r}$.

Consider the possibility of a uniform solution $f_u({\bf r}) = f_0$ of 
Eq.\ (\ref{gl1n}).  Setting  $\nabla^2f = 0$ and using Eqs.\ (\ref{chi}) and 
(\ref{kj}) gives

\begin{equation}\label{flatQ}
{\bf Q}_u = \pm\hat{\bf q}q_u,~~~~~q_u = {k_j\over l_{\beta}\chi^2} = 
{1\over l_{\beta}}\sqrt{-[\mbox{sgn}(\alpha)s^2 + \mbox{sgn}(\beta)\chi^2]}  
\end{equation}
From the definition of ${\bf Q}$, Eq.\ (\ref{Q}), the phase difference 
 over path $r=0$ to $r=d$ is
\begin{equation}\label{flattheta}
\theta({d})_u -\theta({0})_u =  -{2\pi\over\phi_o}\int_0^d d{\bf r}\cdot{\bf A} 
\pm {q}_u {d}.
\end{equation}
Since the constant $q_u$ is real, 
the case with $\mbox{sgn}(\alpha) = 
\mbox{sgn}(\beta) = 1$ is excluded. Further restrictions are given in 
Eq.\ (\ref{flatcond}) below.  Using Eq.\ (\ref{flattheta}), the uniform 
solution has the general form
\begin{equation}\label{flatpsi}
\psi({\bf r})_u = 
\exp\left(-{2\pi i\over\phi_o}\int d{\bf r}\cdot{\bf A}\right)
\left [c_+e^{i{q}_u {r}} + c_-e^{-i{q}_u{r}}\right]
\end{equation}
As will be shown, the uniform solution plays an important role in the general 
solution space. This solution appears, not only at a second order phase 
transition where $|\psi| \rightarrow 0$, but also well below the transition 
point where $|\psi|$ is large.   
 
%==========================================================================

\section{One Dimensional Systems}
\subsection*{Analytic Solution Classification}
In this section a complete set of analytic solutions of the first GL equation 
is derived for one dimensional systems, which in accordance with 
Eq.\ (\ref{divj}) have uniform current density.  In general Eqs.\ (\ref{gl1n}) 
and (\ref{gl2n}) are coupled since ${\bf Q}$ remains multidimensional.  
However in SC microwires,\cite{degennes2} and certain thin films of thickness 
$\le \lambda_{GL}$ of arbitrary widths \cite{tinkham}, the current density is 
approximately uniform.  Consider a one dimensional system with coordinate $x$.  
Approximating the magnitude of the true current density by its mean value over 
all $x$, and  using $J = l_{\alpha} f^2 Q$,  the first GL equation (\ref{gl1n}) 
is
\begin{equation}\label{gl1n1D}
l_{\alpha}^2{d^2f\over dx^2} = \mbox{sgn}(\alpha)f  + 
\mbox{sgn}(\beta) f^3 + J^2f^{-3},
\end{equation}
subject to the boundary condition
\begin{equation}\label{bc1D}
{1\over 2}\left[(a+1)f^{-1}{df\over dx} +(a-1)\left({df\over dx}\right)^{-1} 
{d^2f\over dx^2}\right]_b  = -{1\over b_2} - {1\over b_4}s^2f^2|_b,
\end{equation}
where the subscript $b$ denotes the surface boundary.  At this point the 
rationale for introducing the normalized function $f$ and current parameter 
$J$ is evident.  Since one may use $x/l_{\alpha}$ as a normalized coordinate, 
the only parameter in Eq.\ (\ref{gl1n1D}) is $J$.  The unitless current 
parameter $J$, together with sgn$(\alpha) = \pm 1$, and sgn$(\beta) = \pm 1$ 
completely characterize the solutions of Eq.\ (\ref{gl1n1D}).

When dealing with current carrying states, it is useful to introduce 
$N(x) = f^2(x)$. Equations\ (\ref{gl1n1D}) and (\ref{bc1D}) assume the forms

\begin{equation}\label{d2Ndx2}
Nl_{\alpha}^2{d^2N\over dx^2} = 2J^2 +2 N^2[\mbox{sgn}(\beta)N + 
\mbox{sgn}(\alpha)] + {1\over 2}\left(l_{\alpha}{dN\over dx}\right)^2,
\end{equation} 

and 

\begin{equation}\label{bc1DN}
{1\over 2}\left[{1\over 2}(a+1)N^{-1}{dN\over dx} + 2(a-1)[J^2 + 
N^2(\mbox{sgn}(\beta)N + 
\mbox{sgn}(\alpha))]\left(l_\alpha^2N{dN\over dx}\right)^{-1}\right]_b  = 
-{1\over b_2} - {1\over b_4}s^2N|_b,
\end{equation}
which will be used below. 

Integrating Eq.\ (\ref{gl1n1D}), assuming $J$ constant gives
 
\begin{equation}\label{dfdxC}
2\left(l_{\alpha}{df\over dx}\right)^2 = \mbox{sgn}(\beta)f^4 + 
2\mbox{sgn}(\alpha)f^2 - 2J^2f^{-2} + C,
\end{equation}
where $C$ is a constant of integration.   All solutions of Eq.\ (\ref{dfdxC}) 
may be expressed in terms of Jacobian elliptic functions pq$[u(x)|m]$, which 
are periodic in the argument $u(x)$. In general when $J \ne 0$, determination 
of these functions requires the roots of a cubic equation in $f^2$.  However, 
the process is considerably simplified by dividing the solutions into two 
classes: 1. Functions with at least one positive finite extremum at $x = x_0$, 
at which point $C$ is evaluated.  2. Functions with $f(x = x_w) = 0$, at which 
point $C$ is evaluated.  In the first class only roots of a quadratic equation 
are required.  For the second class only cases with $J = 0$ are possible, and 
these may also be resolved by a quadratic. 
%----------------------------------------------------------------------------
\subsection*{Solutions with Finite, Positive Extrema}
If $f$ has at least one finite extremum point within each period, the constant 
$C$ may be evaluated at an extremum point $x_0$, where $df/dx = 0$.  Thus, we 
write Eq.\ (\ref{dfdxC}) in the form 

\begin{equation}\label{dfdx}
2f^2\left(l_{\alpha}{df\over dx}\right)^2 = 
(f^2 - f_0^2)\{\mbox{sgn}(\beta)f^4 + 
[\mbox{sgn}(\beta)f_0^2 + 2\mbox{sgn}(\alpha)]f^2 + 2J^2f_0^{-2}\},
\end{equation}
where $f_0 = f(x_0)$ is an extremum to be determined from boundary conditions. 
 A second extremum $f_1^2 = f^2(x_1)$, given in terms of $f_0$ by setting the 
 factor $\{...\} = 0$, may exist for both real and imaginary $f_1$. 

Setting $N(x) = f^2(x)$, Eq.\ (\ref{dfdx}) assumes the form
\begin{equation}\label{dNdx}
\left(l_{\alpha}{dN\over dx}\right)^2 = 2(N - N_0)\{\mbox{sgn}(\beta)N^2 + 
[\mbox{sgn}(\beta)N_0 + 2\mbox{sgn}(\alpha)]N + 2J^2N_0^{-1}\},
\end{equation}

Since the solutions are referenced to the extremum value $N_0 = N(x_0)$, 
it follows from Eq.\ (\ref{d2Ndx2}) that the sign of the parameter $b_0$, 
given by
\begin{equation}\label{d2Ndx20}
b_0 = l_{\alpha}^2 {d^2N\over dx^2}|_{x_0} = 
{2\over N_0}[ J^2 - C_0 ],~~\mbox{with}~~ C_0  = 
-N_0^2[\mbox{sgn}(\beta)N_0 + \mbox{sgn}(\alpha)],
\end{equation}
determines whether $N_0$ is a maximum or a minimum.
 
For mathematical expediency, it is useful to make the following variable shift:

\begin{equation}\label{defg}
N(x) = N_0 + \mbox{sgn}(b_0)\nu^2(x).
\end{equation}
Equation (\ref{dNdx}) yields

\begin{equation}\label{dgdx}
2\left(l_{\alpha}{d\nu\over dx}\right)^2 = 
\mbox{sgn}(b_0)\mbox{sgn}(\beta)(\nu^2 - r_+)(\nu^2 - r_-),
\end{equation}
which has roots
\begin{equation}\label{roots}
r_{\pm} = \gamma \pm \Delta ,
\end{equation}
with
\[\gamma = -\mbox{sgn}(b_0)\left[{3\over 2}N_0 + 
\mbox{sgn}(\alpha)\mbox{sgn}(\beta)\right],~~~~\Delta = 
\sqrt{\gamma^2 -b_0\mbox{sgn}(\beta)} = 
\sqrt{{2\over N_0}[C_1 - J^2\mbox{sgn}(\beta)]},~
\]
\[C_0  = -N_0^2[\mbox{sgn}(\beta)N_0 + \mbox{sgn}(\alpha)],~~~~ C_1 = 
{N_0\over 2}\left[{N_0\over 2} + \mbox{sgn}(\alpha)\mbox{sgn}(\beta) \right]^2. 
\]
For $J = 0$, one must use the magnitude of $C_1$ to obtain the correct result 
$\Delta \ge 0$.  The solutions of Eq. (\ref{dgdx}) are the various elliptic 
function combinations, listed in (\ref{solJ}) and (\ref{solJ0}), whose specific 
form is dictated by the characteristics of the roots shown in (\ref{roottable}).
\\[.3cm] 
{\bf Roots}: The array below shows the root ordering for all sign combinations 
that yield real $N(x) \ge 0$ solutions.  The dashed line separates the positive 
$\beta$ cases from the negative $\beta$ cases.\\ 
\begin{equation}\label{roottable}
\begin{array}{lrrrlll}
\mbox{Case}&\mbox{sgn}(\alpha)&\mbox{sgn}(\beta)&\mbox{sgn}(b_0) 
&\mbox{Root Order} & N_0~ \mbox{Range} & ~~~J^2~ \mbox{Range}\\[.3cm]
\mbox{A}&-1~&1~&-1~& r_+  > 0,~~ r_- < 0~~ & 0 \le N_0 \le 1 
&~~~J^2 \le C_0 \le C_1\\
\mbox{B} &-1~&1~&1~& r_+  >  r_- > 0 & 0 \le N_0 \le 2/3 
&~~~ C_0 \le J^2 \le C_1\\    
\mbox{C} &-1~&1~&1~& r_-  <  r_+ < 0 & N_0 \ge 2/3 & ~~~C_0 \le J^2 \le C_1\\
\mbox{D}_{\pm} &\pm 1~&1~&1~& r_- = r_+^* & 0 \le N_0 < \infty 
& ~~~ C_0 \le C_1 \le J^2\\
\mbox{E} &1~&1~&1~&  r_-  <  r_+ < 0 & 0 \le N_0 <\infty 
& ~~~ J^2 \le C_1\\[.1cm]   
--&----&---&---&--------&-------&-----\\ [.1cm]
\mbox{F} &1~&-1~&1~& r_+  > 0,~~ r_- < 0~~  & 0 \le N_0 <\infty 
& ~~~ J^2 \ge C_0\\
\mbox{G} &-1~&-1~&1~& r_+  > 0,~~ r_- < 0~~  & 0\le N_0 <\infty 
& ~~~ J^2 \ge C_0\\
\mbox{H} &1~&-1~&-1~&  r_+  >  r_- > 0 & N_0 \ge 1 & ~~~ J^2 \le C_0\\
\mbox{I} &-1~&-1~&-1~&  r_+  >  r_- > 0 & 0 \le N_0 <\infty & ~~~ J^2 \le C_0
\end{array}
\end{equation}

\noindent
{\bf Solutions with} ${\bf J \neq 0}$:  The function $\nu(x)$ are Jacobian 
elliptic functions of the form pq$[u(x)|m]$, shown in (\ref{solJ})\\
\begin{equation}\label{solJ}
\begin{array}{ccccc}
\mbox{Cases}& \nu(x) & u(x) & m  \\[.3cm]

\mbox{A, F, G} & \displaystyle{\sqrt{{r_+|r_-|\over r_+ + 
|r_-|}}}~\mbox{sd}(u|m) & \displaystyle{{x - 
x_0\over l_{\alpha}}\sqrt{{ r_+ + |r_-|\over 2}}} 
&\displaystyle{ {r_+ \over  r_+ + |r_-|}}\\[.3cm] 

\mbox{B, H, I} & \sqrt{r_-}~\mbox{sn}(u|m) 
& \displaystyle {{x - x_0\over l_{\alpha}}\sqrt{{ r_+ \over 2}}} 
& \displaystyle{{r_- \over  r_+}}\\[.3cm]

\mbox{C, E} & \sqrt{|r_+|}~\mbox{sc}(u|m) 
& \displaystyle{{x - x_0\over l_{\alpha}}\sqrt{{|r_-|\over 2}}} 
& \displaystyle{1 - {|r_+|\over |r_-|}}\\[.3cm]

\mbox{D}_{\pm} & \sqrt{|r_+|}~\mbox{sc}(u|m)\mbox{dn}(u|m) 
& \displaystyle{{x - x_0\over l_{\alpha}}\sqrt{{|r_+|\over 2}}} 
& \displaystyle{{1\over 2}\left[1 + {Re(r_+) \over  |r_+|}\right]},
\end{array}
\end{equation}
where $|r_\pm| = - r_\pm$ if $r_\pm < 0$ for real roots, and  
$|r_\pm| = \sqrt{r_+r_-} = \sqrt{b_0}$ for the complex root case D.   The 
function $N(x) = f^2(x)$ is given by Eq.\ (\ref{defg}).  Solutions for a 
conventional SC are cases A-C and D with sgn$(\alpha) = -1$ and 
sgn$(\beta) = 1$, which are the same as those given in Ref.\onlinecite{ammann}.
  Solutions for the remaining 6 anomalous cases fall into the same solution 
  classes, however $\gamma, b_0, C_0, C_1$ and $\Delta$ are different 
  functions of $N_0$, due to different values of sgn$(\alpha)$ and 
  sgn$(\beta)$.\\[.5cm]

\noindent
{\bf Solutions with} ${\bf J =  0}$: It follows from Eq.\ (\ref{dfdx}), 
with $J = 0$, that $f(x)$ may be expressed directly in the form $f(x) = 
f_0 pq(u|m)$. \\

\begin{equation}\label{solJ0}
\begin{array}{cccccc}
\mbox{Case}& f(x)/f_0 & u(x) & m & N_0~\mbox{Range} & \mbox{sgn}(b_0)\\[.3cm]

\mbox{A} &{\mbox{cn}\over \mbox{dn}} 
& \displaystyle{{x - x_0\over l_{\alpha}}\sqrt{1-{N_0\over 2}}} 
& \displaystyle{{N_0\over 2-N_0}} & 0 \le N_0 \le 1 & -1\\[.3cm]

\mbox{C1} &{\mbox{dn}\over \mbox{cn}} 
& \displaystyle{ {x - x_0\over l_{\alpha}}\sqrt{{N_0\over 2}}} 
& \displaystyle{{2 -N_0\over N_0}} &1 \le N_0 \le 2 & 1\\[.3cm]

\mbox{C2} &{1\over\mbox{cn}} 
& \displaystyle{{x - x_0\over l_{\alpha}}}\sqrt{N_0 -1} 
&\displaystyle{{0.5N_0-1\over N_0 - 1}} & N_0 \ge 2  & 1\\[.3cm]

\mbox{E} &{1\over\mbox{cn}} 
& \displaystyle{{x - x_0\over l_{\alpha}}}\sqrt{N_0 + 1} 
&\displaystyle{{0.5N_0 + 1\over N_0 + 1}} &0 \le N_0 <\infty & 1\\ [.3cm]

---&-----&------&--------&------&---\\ [.3cm]

\mbox{F} &{1\over\mbox{dn}} 
& \displaystyle{{x - x_0\over l_{\alpha}}\sqrt{1-{N_0\over 2}}} 
& \displaystyle{{1-N_0\over 1-0.5N_0}} & 0 \le N_0 \le 1 & 1\\[.3cm]

\mbox{H1} &\mbox{dn} 
&\displaystyle{{x - x_0\over l_{\alpha}}\sqrt{{N_0\over 2}}} 
& \displaystyle{{N_0-1\over 0.5N_0}} & 1 \le N_0 \le 2& -1\\[.3cm]
 
\mbox{H2} &\mbox{cn} 
& \displaystyle{{x - x_0\over l_{\alpha}}}\sqrt{N_0- 1} 
&\displaystyle{{0.5N_0\over N_0 -1}} & N_0 \ge 2 & -1\\[.3cm]

\mbox{I} &\mbox{cn} 
& \displaystyle{{x - x_0\over l_{\alpha}}}\sqrt{N_0 + 1} 
& \displaystyle{{0.5N_0\over N_0 + 1}} & 0 \le N_0 <\infty &-1 
\end{array}
\end{equation}
The zero current solutions in (\ref{solJ0}) may also be obtained from 
Eq.\ (\ref{solJ}).  The dashed line in (\ref{solJ0}) separates the 
sgn$(\beta) = 1$ cases from the sgn$(\beta) = -1$ cases. The latter cases 
are possible only when superconductivity of the surface of the specimen is 
strongly enhanced through the boundary condition by introducing a significant 
quantity of pair interactions.  For sgn$(b_0) = +1, (-1)$,  the function $f_0 = 
f(x_0)$ is a minimum(maximum), respectively. 

We emphasize that $f^2 = N$ is employed only for mathematical convenience. 
It is particularly useful in numerical computation when one needs the precise 
range of $N_0$ which determines the parameter $m$ in (\ref{solJ0}).  However, 
the function $\chi = sf$ is the normalized order parameter of physical interest,
 with $|\psi |^2 = n_s^*\chi^2$ the actual superconducting pair density. 
  Numerous treatises based on the GL model of superconductivity use the 
  parameter $f$ to illustrate the behavior of superconductors. In general, 
  this is not a good parameter to use, except as an intermediate mathematical 
  step.   As seen in (\ref{solJ0}) the parameter $f$ can become singular in 
  cases C2, E, H2, and I, and is larger than unity in cases C1 and H1. As shown 
  below, $f \rightarrow \infty$ when $s \rightarrow 0$; whereas  $\chi = sf$ 
  remains finite and well behaved.  Similar comments apply to the use of the 
  current parameter $k_j$ in lieu of $J$.   
  {\it Thus it is $\chi$( or $|\psi |$) and $k_j$ that should generally be used 
  for final illustration of superconducting effects.}

It is interesting and useful to examine special solutions that appear at 
certain points in the $(J, N_0, s)$ parameter space.  At the end of Section II 
we determined the complex order parameter for the uniform solution $\chi_u$. 
Now we determine $\chi_u$ from the boundary conditions.  Setting 
$dN/dx = d^2N/dx^2 = 0$ for all values of $x$, Eq.\ (\ref{d2Ndx2}) requires 
that $J^2 = C_0$, i.e. $b_0 = 0$. It follows, using Eq.\ (\ref{kj}),
 $\chi = s\sqrt{N}$, and the boundary condition (\ref{bc1D}) with $a = 1$, 
 that the $(\chi_u, s_u)$ coordinates of the uniform solution $\chi_u$ are
\begin{equation}\label{sflat} 
\chi_u = s_u(0) = \sqrt{-b_4/b_2},~~~~~~s_u(k_j) = 
s_u(0)\sqrt{-\mbox{sgn}(\alpha)[\mbox{sgn}(\beta) + (k_j/\chi_u^3)^2]}. 
\end{equation}
Since $\chi/s = \sqrt{N}$, the argument of the root in Eq.\ (\ref{sflat}) is 
$1/N_u$.  For the uniform solution to exist for all $x$, the sgn$(b_2)$ = 
-sgn$(b_4)$, and the order parameter $\chi_u$ is independent of the current 
parameter $k_j$; whereas $s_u$ depends on $k_j$.  Because $s$ is real and 
positive, Eq.\ (\ref{sflat}) requires the $[\mbox{sgn}(\alpha), 
\mbox{sgn}(\beta)]$ combinations and $k_j$ ranges shown in 
array (\ref{flatcond}).

\begin{equation}\label{flatcond}
\begin{array}{crrc}
\mbox{Cases}& \mbox{sgn}(\alpha)  & \mbox{sgn}(\beta)  
& k_j \mbox{Range} \\[.3cm]
\mbox{A}|\mbox{B, C} & -1 &1 & k_j \ge 0\\[.3cm]

\mbox{F}|\mbox{H} &~~~1 &-1 & ~~~0 \le k_j \le \chi^3\\[.3cm]

\mbox{G}|\mbox{I} &-1 &-1 & k_j \ge \chi^3.
\end{array}
\end{equation} 
Applying the uniform solution condition $b_0 = 0$ to the elliptic function 
solutions in (\ref{solJ}) gives $N(x) = N_0$, with $m = 0$, at the boundaries 
of the solution case domains indicated in Eq.\ (\ref{flatcond}) by a vertical 
line.  When $k_j \ne 0$, the uniform solution point $s_u$ approaches infinity 
in both the large and small limits of $\chi$, corresponding, respectively, to 
weak and strong surface interaction limits.

Since $s(t) = \sqrt{|\alpha(t)|/|\beta(t)|}$ by definition, 
Eq.\ (\ref{sflat}) gives the relationship at $t = t_u$ between the volume 
parameter ratio $\alpha/ \beta$ and the surfaces parameter ratio $b_4, b_2$ 
as a function of the current parameter $k_j$.  For $k_j = 0$, which requires 
that $\mbox{sgn}(\alpha) = - \mbox{sgn}(\beta)$, it follows that
\begin{equation}\label{volsurf}
{\alpha(t_u)\over \beta(t_u)} =  -\chi_u^2(t_u) = {b_4\over b_2},
\end{equation} 
where $t_u$ is the normalized temperature at which $s(t_u) = s_u$.
For a bulk SC with $k_j = 0$, the ratio $\alpha(0)/ \beta(0) = -1$ and 
$\chi(0) = 1$. In contrast, the ratio $\alpha(t_u)/ \beta(t_u)$ given by 
Eq.\ (\ref{volsurf}) is generally not equal to $-1$.     

The uniform solutions and other special solutions follow from the general 
solutions listed in Eqs.\ (\ref{solJ}) and (\ref{solJ0}). For $J = 0$ they are 
listed in Eq.\ (\ref{special}) below.

\begin{equation}\label{special}
\begin{array}{ccccc}
\mbox{Case}& f(x)/f_0  & m & N_0 \\[.3cm]

\mbox{A, I} &~~~~ \cos\left(\displaystyle{{x - x_0\over l_{\beta}}s}\right)  
&~~~ 0 &~~~\rightarrow 0\\[.3cm]

\mbox{E, F} &~~~~ \cosh\left(\displaystyle{{x - x_0\over l_{\beta}}s}\right) 
&~~~ 1 &~~~\rightarrow 0\\[.3cm]

\mbox{A}|\mbox{C1} &~~~~ 1  &~~~ 1 &~~~1\\[.3cm]

\mbox{F}|\mbox{H1} &~~~~ 1  &~~~ 0 &~~~1\\[.3cm]

\mbox{C1}|\mbox{C2, D}_- 
&~~~~ \mbox{sec}\left(\displaystyle{{x - x_0\over l_{\beta}}s}\right) 
&~~~ 0 &~~~2\\[.3cm]

\mbox{H1}|\mbox{H2} 
&~~~~\mbox{sech}\left(\displaystyle{{x - x_0\over l_{\beta}}s}\right) 
&~~~ 1 &~~~2\\[.3cm]

\mbox{C2}|\mbox{E, H2}|\mbox{I} 
&~~~~  \mbox{nc}\left[\displaystyle{{x - x_0\over l_{\beta}}\chi(0)}\right]  
&~~~ {1\over2} &~~~\infty         
\end{array}
\end{equation}
The vertical line separating cases indicates that the special solution occurs 
at the boundary between the case solution domains. The solutions with 
$N_0 \rightarrow 0$ are second order phase transition solutions which satisfy 
the linearized form of Eq.\ (\ref{gl1n1D}); those with $N_0 = 1$ are the $x$ 
independent uniform solutions; those with $N_0 = 2$ satisfy the nonlinear form 
of Eq.\ (\ref{gl1n1D}); and those with $N_0 \rightarrow \infty$ have $s = 0$ 
and represent the transition between Jacobian elliptic functions with and 
without inflection.
%--------------------------------------------------------------------------
\subsection*{Solutions for $J = 0$ with $f(x = x_w) = 0$}
When $J = 0$, Eq.\ (\ref{dfdxC}) assumes the form
\begin{equation}\label{dfdxCr}
2\left(l_{\alpha}{df\over dx}\right)^2 = \mbox{sgn}(\beta)f^4 + 
2\,\mbox{sgn}(\alpha)f^2 + C = \mbox{sgn}(\beta)(f^2 -r_+)(f^2 -r_-),
\end{equation}
where
\begin{equation}\label{wetroots}
r_\pm = -\mbox{sgn}(\alpha)\mbox{sgn}(\beta) \pm \sqrt{1-C\,\mbox{sgn}(\beta)}.
\end{equation}
Although the constant $C$ is arbitrary, the only solutions of interest that 
were not included in solutions with finite extrema are those for which $f$ is 
zero at some point $x_w$, with $df/dx|x_w \ne 0$. In such cases, the constant 
$C$ is  
\begin{equation}\label{C} 
C = 2\left(l_{\alpha}{df(x_w)\over dx}\right)^2 \ge 0.
\end{equation}
The root ordering and corresponding solutions are given in 
Eq.\ (\ref{wetroottable}) and (\ref{wetsol}) below.
\begin{equation}\label{wetroottable}
\begin{array}{lrrlc}
\mbox{Case}&\mbox{sgn}(\alpha)&~~~\mbox{sgn}(\beta) &\mbox{Root Order}
& \mbox{C Range}\\[.3cm]
\mbox{A'}&-1~~&1~~~~& r_+  >  r_- > 0& C < 1\\[.1cm]  
\mbox{D'} &\pm 1~~&1~~~~& r_- = r_+^*& C > 1 \\[.1cm]  
\mbox{E'} &1~&1~~~~&  r_-  <  r_+ < 0& C < 1 \\[.1cm]                           
\mbox{F', I'} &\pm 1~~&-1~~~~& r_+  > 0,~~ r_- < 0&~~~  0 \le C < \infty
\end{array}
\end{equation}

\begin{equation}\label{wetsol}
\begin{array}{lccl}
\mbox{Cases}&~~ f(x)~~ &~~ u(x)~~~~~ &~~~~ m  \\[.3cm]

\mbox{A'} & \sqrt{r_-}~\mbox{sn}(u|m) 
& \displaystyle {{x - x_1\over l_{\alpha}}\sqrt{{ r_+ \over 2}}} 
& \displaystyle{{r_- \over  r_+}}\\[.3cm] 

\mbox{D'} & \sqrt{|r_+|}~\mbox{sc}(u|m)\mbox{dn}(u|m) 
& \displaystyle{{x - x_1\over l_{\alpha}}\sqrt{{|r_+|\over 2}}} 
& \displaystyle{{1\over 2}\left[1 + {Re(r_+) \over  |r_+|}\right]}\\[.3cm] 

\mbox{E'} & \sqrt{|r_+|}~\mbox{sc}(u|m) 
& \displaystyle{{x - x_1\over l_{\alpha}}\sqrt{{|r_-|\over 2}}} 
& \displaystyle{1 - {|r_+|\over |r_-|}}\\[.3cm] 

\mbox{F', I'} 
& \displaystyle{\sqrt{{r_+|r_-|\over r_+ + |r_-|}}}~\mbox{sd}(u|m) 
& \displaystyle{{x - x_1\over l_{\alpha}}\sqrt{{ r_+ + |r_-|\over 2}}}~~~~ 
&\displaystyle{ {r_+ \over  r_+ + |r_-|}},
\end{array}
\end{equation}
where $r_\pm$ are given by Eq.\ (\ref{wetroots}), with $|r_\pm| = - r_\pm$ if 
$r_\pm < 0$ for real roots, and  $|r_\pm| = \sqrt{r_+r_-} = \sqrt{C}$ for 
case D'.  In the application considered next, some of the functions in 
Eq.\ (\ref{wetsol}) are used to construct surface ``pre-wetting'' 
solutions\cite{montevecchi} which are zero at a point $x_1$ near a surface. 
Since $f \ge 0$ by definition, only that part of the elliptic function that 
is positive represents a physical solution. The slope of $f^2(x)$
at $x = x_1$ is zero, thus establishing a smooth
transition to the region with $f = 0$.

%=============================================================================
\section{Application: Plane Slab Sample}

The structure used in this section is a symmetric plane slab of thickness $d$, 
with the origin $x = 0$ at its center.  First we consider the zero current slab 
body solutions with $N_0 = N(0) = f^2(0) \ne 0$, then the corresponding current 
supporting solutions, followed by the currentless surface ``pre-wetting'' 
solutions with $f(x_w) = 0$. 
%----------------------------------------------------------------------
\subsection*{Currentless Slab Solutions}

 For $J = 0$ there are eight solutions listed in (\ref{solJ0}).  Since $N(0)$ 
 may become singular, we replace it with the well behaved normalized order 
 parameter $\chi^2(0) = s^2N(0)$, and use $ l_{\beta} = sl_{\alpha}$.  Applying 
 the boundary condition (\ref{bc1D}) at $x = 0.5 d$, setting $x_0 = 0$, and 
 defining $u_s = u(x = 0.5d)$, leads to the boundary condition   
\begin{equation}\label{bcchi}
a\Omega_+ - \Omega_-  +  {d\over b_4}\chi^2(0.5d) = -{d\over b_2},
\end{equation}
where
\[\Omega_{\pm} =
 {d\over 2}\left[\left({d\chi\over dx}\right)^{-1}
 {d^2\chi\over dx^2} \pm \chi^{-1}{d\chi\over dx}\right]_{0.5d}.\]
When $a = 1~(c = \epsilon)$, it is seen in Eq.\ (\ref{bcchi}) that only the 
logarithmic derivative $\chi^{-1}d\chi/dx$ is specified at the boundary.  
As mentioned, and as will be shown, $a = 1$ is the important case permitting 
second order phase transitions. 

With considerable manipulation of the Jacobian elliptic functions, one may 
derive the following forms for $\Omega_{\pm}$.

\begin{equation}\label{Omegapm}
\begin{array}{cc}
\mbox{Case}&\Omega_{\pm} \\[.3cm]
\mbox{A}&~~~~\displaystyle{{2u_s\over\mbox{sn}(2u_s)}}
\left[1 + {1\over 2}(1 \pm 1)[\mbox{cn}(2u_s) - \mbox{dn}(2u_s)]\right]\\[.4cm]

\mbox{C1}&~~~~\displaystyle{{2u_s\over\mbox{sn}(2u_s)}}
\left[1 - {1\over 2}(1 \pm 1)[\mbox{cn}(2u_s) - \mbox{dn}(2u_s)]\right]\\[.4cm]

\mbox{C2,E}&~~~~\displaystyle{{2u_s\over\mbox{sn}(2u_s)}}
\left[\mbox{dn}(2u_s) + {1\over 2}(1 \pm 1)[1-\mbox{cn}(2u_s)]\right]\\[.4cm]

\mbox{F}&~~~~\displaystyle{{2u_s\over\mbox{sn}(2u_s)}}
\left[\mbox{cn}(2u_s) + {1\over 2}(1 \pm 1)[1-\mbox{dn}(2u_s)]\right]\\[.4cm]

\mbox{H1}&~~~~\displaystyle{{2u_s\over\mbox{sn}(2u_s)}}
\left[\mbox{cn}(2u_s) -{1\over 2}(1 \pm 1)[1-\mbox{dn}(2u_s)]\right]\\[.4cm]

\mbox{H2,I}&~~~~\displaystyle{{2u_s\over\mbox{sn}(2u_s)}}
\left[\mbox{dn}(2u_s) - {1\over 2}(1 \pm 1)[1-\mbox{cn}(2u_s)]\right]\\[.4cm]

\end{array}
\end{equation} 
The argument $u_s = u(x = 0.5d)$ and parameter $m$ are easily transformed in 
(\ref{solJ0}) using $\chi^2(0) = s^2N(0)$.  The boundary conditions listed by 
Eqs.\ (\ref{bcchi}) contain the set $\{\chi(0), s(t), d/l_{\beta}, d/b_2 , 
d/b_4 \}$.   Specifying the three scaled lengths, the boundary equations give 
the normalized order parameter $\chi(0)$ at the slab center as a function of 
$s$.  In the boundary condition (\ref{bcchi})  the cases C and H in  
(\ref{solJ}) split into subcases, defined in (\ref{solJ0}). The sgn($\alpha$) 
and sgn($\beta$) remain those of the parent case.  The cases in (\ref{bcchi}) 
with sgn$(b_0) = -1$ have superconductivity at the surface reduced from that 
in the sample bulk; whereas cases with sgn$(b_0) = 1$ have enhanced surface 
superconductivity.

An alternative form of the boundary condition Eq.\ (\ref{bc1D}) is obtained 
by integrating the first GL Eq.\ (\ref{gl1n1D}) across the slab. Noting that 
$df(x)/dx = -df(-x)/dx$ and $f(x) = f(-x)$, one obtains for the case $a = 1$ 
the relation

\begin{equation}\label{bcint}
{1\over b_2} + {1\over b_4}s^2f^2(0.5d) = -{1\over l_{\alpha}^2f(0.5d)}
\int_0^{0.5d} dx[\mbox{sgn}(\alpha)f(x) + \mbox{sgn}(\beta) f^3 + J^2f^{-3}].
\end{equation}
Since Eq.\ (\ref{bcint}) is exact within the GL model, it is difficult to 
reconcile it with a similar, but distinctly different expression involving the 
density of states given by de Gennes\cite{degennes}, who uses the same 
arguments to derive the GL equations.

Up to this point, all relations follow directly or indirectly from the free 
energy (\ref{G1}) with the models for $U$, Eq.\ (\ref{Upot}), and $\Lambda$, 
Eq.\ (\ref{Lambdas}).  No approximations, assumptions, or other conjectures 
have been made.  To solve the boundary Eqs.\ (\ref{bcchi}) we  assume that 
$l_{\beta}$ is approximately temperature independent.   Comparing the surface 
energy density with the volume condensation energy density, one notes that 
$1/b_2(t)$ and $|\alpha(t)| \propto s^2(t)$ are both coefficients which scale 
$|\psi|^2$, and that $1/b_4$ and $|\beta|$ both scale $|\psi|^4$.  Hence, by 
analogy, we conjecture that $1/b_2(t)$ may be a function of $s^2(t)$ and $b_4$ 
is approximately independent of temperature.  Since we have no model for 
$b_2(t)$ we will neglect the temperature dependence and use  
$b_2(t) \approx b_2$, where $b_2$ is a controllable external parameter. 
[Assuming that $1/b_2(t) \propto s^2$ leads to unreasonable results.]

The parameters $b_2$, and $b_4$ cannot be chosen arbitrarily. The free energy 
must be negative for a SC state to exist, which limits the range of $b_4$, and 
the boundary condition at a second order phase transition limits the value and 
sign of $b_2$.   From Eq.\ (\ref{GminMod}), the normalized, minimum free 
energy for a symmetric slab is 
\begin{equation}\label{Gamin}
\Gamma_{min}(s) = (2\pi)^2{\Delta G_{min}\over K^*_d} =  
-{1\over 2}\left({d\over l_{\beta}}\right)^2
\langle \chi^4\rangle\mbox{sgn}(\beta) - 
{d\over b_4}\chi^4(0.5d) - 2(a-1)\chi^2(0.5d)\Omega_-,
\end{equation}
where

\[K^*_d = n_s^*V\left({1\over 2m^*}\right)\left({h\over d}\right)^2,~~~~~
\langle \chi^4\rangle = {2\over d} \int_0^{0.5d} dx\chi^4(x).\]
The normalizing term $K^*_d$ is the total kinetic energy of a non-interacting 
gas of SC pairs in volume $V =\mbox{ area}\times d$, with wavelength $d$ at 
$t = 0$.  The entire slab is in a SC state providing that the parameter 
$\Gamma_{min}(s) < 0$.  It is clear that if $a = 1~(c = \epsilon)$ the 
inequality  $\Gamma_{min} < 0$ can be satisfied when $\beta > 0$ for all $b_4 > 
0 $, or for $b_4 < 0$ in a restricted range. When $\beta <  0$ then  $b_4$ must 
be positive, but not too large. For $a \ne 1$ the key factor determining 
whether a state is superconducting is the sgn$(\Gamma_{min})$.  States that 
exhibit a second order phase transitions\,(SOPT) are defined by the zero limit 
entropy difference between S and N states, 
\begin{equation}\label{sopt}
\lim_{\chi\rightarrow 0}{\partial\Gamma_{min}\over \partial t} =  
{\partial s\over \partial t} \lim_{\chi\rightarrow 0}
{\partial\Gamma_{min}\over \partial s} = 0.
\end{equation}  
It can be shown that all states with $a \ge 1$ are potentially superconducting 
states, and in restricted $s(t)$ value domains states with $a \le 1$ may be 
superconducting with a first order phase transition.  However only states with 
$a = 1$ have second order phase transitions, as defined by Eq.\ (\ref{sopt}). 
  In this paper we focus on the SOPT states and states that are continuations 
  of the SOPT states corresponding to the case $a = 1$. 

When $J = 0$ the uniform solutions are $\chi_u = s_u = \sqrt{- b_4/b_2}$, 
which may exist for cases A, C1, F, and H1.  The normalized energy for $a = 1$ 
has the value
\begin{equation}\label{Gaminflat}
\Gamma_{min} =  
-{1\over 2}\left({b_4\over b_2}\right)^2
\left[\left({d\over l_{\beta}}\right)^2\mbox{sgn}(\beta) + 
{2d\over b_4}\right],
\end{equation}
This exact analytical result for the uniform solution gives a simple check on 
the numerical results.

The point where $\chi(x) \rightarrow 0$ is interpreted as a second order phase 
transition point, and the corresponding $s$-value is obtained from 
Eq.\ (\ref{bcchi}) by letting $\chi(0)$ approach zero.  As can be seen from 
Eq.\ (\ref{solJ0}) cases A and I give rise to a solution with maximum order 
parameter, while solutions E and F with minimum order parameter $\chi(0)$ in 
the center of the slab.  Cases A and I, with sgn$(\alpha ) = -1$, lead to a 
characteristic equation for $s = s_o$ in the limit $\chi(0)\rightarrow 0$. In 
this limit, from (\ref{special}), $\chi(x) = \chi(0)\cos(\zeta x/d)$, and the 
characteristic equation (\ref{bcchi}) with $a = 1$ is

\begin{equation}\label{bcAlin}
\zeta_o\mbox{tan}(0.5\zeta_o) = {d\over b_2} ~~~~\mbox{with}~~~~\zeta_o = 
{d\over l_{\beta}}s_o. 
\end{equation}

Similarly, for $a = 1$ the cases E and F, with sgn$(\alpha ) = 1$, and
 $\chi(x) = \chi(0)\cosh(\zeta x/d)$ gives 

\begin{equation}\label{bcElin}
\zeta_o\mbox{tanh}(0.5\zeta_o) = -{d\over b_2}. 
\end{equation}
 As will be shown below, if $s$ is interpreted as a function of temperature,
  Eq.\ (\ref{bcAlin}) gives rise to a reduction of $t_{cs}$ due to a pulling 
  down of the order parameter at the surfaces of the slab, while 
  Eq.\ (\ref{bcElin}) leads to an enhanced transition temperature 
  $t_{cs} = T_{cs}/T_c$, as previously shown.\cite{simonin,montevecchi}

%----------------------------Fig. Discussion

Fig. \ref{fig1} shows the E solution for $a = 1$ and for $ a = 1 \pm 0.001$ 
near the transition point defined by $\chi \rightarrow 0$.  The solid curves 
are the square of the normalized order parameter $\chi(0, s)\equiv \chi(0)$ 
plotted as a function of the length ratio $s(t) = l_{\beta}/l_{\alpha}(t)$, 
and the dashed curves are the energy parameter $\Gamma_{min}(s)$.  Only the 
curve for $a = 1$ satisfies the entropy condition (\ref{sopt}) for a second 
order phase transition. For a fixed set of parameters, the $a = 1$ states 
always have a lower value of $\Gamma_{min}(s)$ than states with $a \ne 1$ at 
the same value of $s(t)$.

As we go below the second order phase transition point the order parameter 
increases, but the corresponding $s(t)$ value varies in different ways 
depending on the type of solution of the nonlinear equations,  
Eq.\ (\ref{bcchi}).  As can be seen from Eq.\ (\ref{Gamin}), the minimum free 
energy $\Gamma_{min}$ can be positive or negative depending on sgn($\beta$) 
and the value of $b_4$, and implicitly on $b_2$ through $\chi$.  In the 
conventional GL approach $\beta > 0$.  

Figs. \ref{fig2} and \ref{fig3} show the zero current solutions of the 
nonlinear Eqs. (\ref{bcchi}) for $a = 1~(c = \epsilon)$ with the scaled order 
parameter at the slab center $\chi(0)$ (solid lines), and the scaled minimum 
free energy $\Gamma_{min}$ (dashed lines), Eq.\ (\ref{Gamin}).  In 
Figs. \ref{fig2}-\ref{fig10} the ratio $d/|b_2| = 0.01$.  For a second order 
phase transition the sign of $b_2$ is dictated by the linearized 
Eqs.\ (\ref{bcAlin}) and (\ref{bcElin}).  The value of $d/l_{\beta} = 0.5$ in 
Figs. \ref{fig2}-\ref{fig10}, i.e. the value of $|\beta|$ is fixed relative to 
the slab thickness $d$.

Figure \ref{fig2} shows our results for cases A, C1, C2, and E, all for 
$\beta > 0$, obtained from Eq.\ (\ref{bcchi}) for various values of $b_2$ and 
$b_4$.  The normalized energy function $\Gamma_{min}$ is shown by the broken 
lines and $\chi(0)$, the normalized order parameter at the center of the slab, 
defined by Eq.\ (\ref{chi}), is depicted by solid lines.  In this particular 
situation $\Gamma_{min}$ is always negative, but very small; thus negative 
values of $b_4$ could drive the slab into the normal state with 
$\Gamma_{min} > 0$.  The regions over which the different solution types are 
valid are separated by a bar, and as can be seen from the intersection of both 
A and C1 curves.  Depending on the values of $b_2$ and $b_4$, the same type of 
solution can appear in different regions of the $\chi(0)$ versus $s$ diagram.  
Solutions A and C1 meet when $N_0 = 1$, at which point 
$\chi_u = s_u = \sqrt{-b_4/b_2} = 1/\sqrt{5}$ is a constant 
[See Eqs.\ (\ref{sflat})-(\ref{special})], independent of the coordinate $x$.  
As $s$ varies through this point, the slope of $\chi(x = d/2)$ changes sign as 
it goes through zero.

The point where $\chi(0) = 0$ is interpreted as the second order phase 
transition point and for a finite value of $b_2$ it occurs at a positive, 
non-zero value of $s$.  For solution A, the order parameter at the surface of 
the slab is ``pulled down'' with respect to the center of the slab while for 
solution E it is ``pulled up''.  We shall interpret below the A case as a 
decrease of the superconducting transition temperature $T_c$ due to the 
proximity effect caused, for example, by a normal metal deposited on the 
surface of a thin film, while the opposite applies to case E.  The symbol for 
the reference temperature is $T_c$, at which point $\chi(0) = 0$ occurs when 
$1/b_2 = 0$.  For case A, as $s$ increases, $\chi(0) = s\sqrt{N_0}$ increases, 
which is interpreted as a decrease in temperature, away from the phase 
transition point. 

For case E the value of $s$ decreases as the order parameter $\chi$ increases. 
 For $s \rightarrow 0$ the value of $\chi(0)$ remains finite when $1/b_2 < 0$. 
  This happens because $N_0 \equiv N(0) = f^2(0) \rightarrow \infty$ when 
  $s \rightarrow 0$.  For $s = 0$ solution type E and C2 are satisfied by 
  $N_0 \rightarrow \infty$ with Jacobian function parameter $m = 0.5$.  The 
  order parameter function $\chi(x)$ is, however, well behaved for $s = 0$ as 
  seen in (\ref{special}) for the C2$|$E point.  As $\chi(0)$ increases, the 
  function of type E switches to a C2 function and the $s$-value increases, as 
  is the case for the A type solution, moving to lower temperatures, away from 
  the phase transition point\,(lower A branch).  The same interpretation is 
  now applied to the upper curve as we go from type C2 to C1 to the upper A 
  curve.  Thus, if the phase transition point for the E type solution is 
  interpreted as occurring at an enhanced transition temperature $T_{cs} > T_c$, 
  then the point $s = 0$ anchors the temperature $T = T_c$. 

As one goes from the phase transition point along the E, C2, C1,and upper A 
branches, the parameter $N_0 = f^2(0)$ goes from zero to infinity at $s = 0$,
 with $\chi(0)$ finite, for the E branch. For the C2 section the function $N_0$ 
 is also infinite at $s = 0$.  As $s$ and $\chi(0)$ increase, C2 changes into 
 C1 at $N_0 = 2$.  At this point the Jacobian elliptic function becomes a 
 trigonometric function\,[See Eq.\ (\ref{special})] and then changes again into 
 an elliptic function in the C1 region.  At the point where the C1 joins the 
 upper A branch, $b_0 \propto d^2\chi/dx^2|_{x = 0}$ changes sign, and the 
 solution $\chi(x)$ becomes independent of $x$ at the intersection.  In domains 
  C2, and C1 the net slope of $\chi(x)$ at the surface $x = d/2$ is positive, 
  while in the upper A domain it is negative.  

The effect of changing the sign of the surface pair interaction parameter 
$b_4$ is also shown.  When $b_4$ is negative it increases the net slope of 
$\chi(x)$ at the surface, enhancing superconductivity there by lowering the 
total energy of the slab.  For solutions E, C2, and C1 the net slope at the 
slab surface $x = d/2$ is always positive, thus aiding superconductivity in 
the slab, while type A gives rise to a negative slope at $x = d/2$ which has 
an inhibiting effect on the superconducting state in the slab.  As mentioned,
 where solutions A and C1 meet, the net slope of the order parameter is zero 
 and $\chi(x)$ is constant, independent of $x$.

Figure \ref{fig3} depicts solutions corresponding to Fig. \ref{fig2} with 
$a = 1$, but with $\beta < 0$ for solutions F, H1, H2, and I.  In order for 
these solutions to be superconducting the total minimized free energy 
difference $\Gamma_{min}$, Eq.\ (\ref{Gamin}), must be negative.   
Equation (\ref{Gamin}) with $a = 1$ shows that this can be achieved only if 
$b_4 > 0$ with the surface term $(d/b_4)\chi^4(0.5d)$ larger than the volume 
term $0.5(d/l_{\beta})^2\langle\chi^4\rangle$.  As for $\beta > 0$ cases, the 
surface parameter $b_2$ determines the point at which the second order phase 
transition occurs.  Again, the results of Fig. \ref{fig3} are exact solutions 
of our generalized GL equations and are completely independent of the 
temperature dependence of $s(t)$.  Although the lower curve for solution type I 
can be interpreted in the same way as the lower A-type curve in Fig.\ref{fig2} 
was interpreted, the rest of Fig. \ref{fig3}  has some notable differences.  
The phase transition point $\chi(0) = 0$ on branch F can be interpreted as an 
enhancement of the transition point above the reference temperature $T_c$, 
defined by $s(t = 1) = 0$, because the order parameter at the surface is
 ``pulled up'' at $x = d/2$.  Where F joins branch H1, the order parameter 
 $\chi = \sqrt{-b_4/b_2} = 1/5$ is independent of $x$; thus the slope at the 
 surface changes sign at this point and H1 becomes a ``pull down'' case before 
 the reference temperature at $s = 0$ is reached.  As the order parameter 
 $\chi(0)$ increases from zero, its value is finite at $s = 0$ as it goes from 
 F type to H1 and H2 and then to a different upper branch of the I solution.  
 Along the H1, H2 and I curves, the order parameter $\chi(x = d/2)$ is always 
 ``pulled down'' with respect to $\chi(0)$. At the point $\chi(0) = \sqrt{2}s$, 
 where H1 and H2 join, $\chi(x)$ becomes a hyperbolic function, 
 (\ref{special}), and at $s = 0$, where H2 and I join, the value of 
 $N_0 = f^2(0) \rightarrow \infty$.  For the given parameter values, all of 
 the solutions are superconducting, which depend on a positive $b_4$ of the 
 pair interaction parameter near the surface.

Although the F solution is the only one with the order parameter ``pulled up'' 
at the surface, all solutions are superconducting. In particular, the H1 and H2 
solutions are ``pull down'' cases existing above our reference point $T_c$ 
defined by $s(T = T_c) = 0$.  Thus enhanced superconductivity can exist for 
$\beta < 0$ as long as the surface pair interaction parameter $1/b_4$ is large 
enough to overcome the unfavorable effect due to sgn$(\beta) = -1$ in the slab 
volume.  This is the case for SNS or SS'S junctions where the N or S' regions 
are embedded by a strong superconductor S.

The results depicted in Figs. \ref{fig2} and \ref{fig3}, for the order 
parameter $\chi(x)$, defined by Eq.\ (\ref{chi}) as a function of the length 
ratio $s(t)$, defined by Eq.\ (\ref{s}), with both $\chi(x)$ and $s(t)$ 
positive quantities by definition, are exact solutions of our generalized, 
nonlinear equations for a slab with $k_j = 0$.  They do not depend on, or show 
how, $s(t)$ is related to the normalized temperature $t = T/T_c$, which is not 
included explicitly in the GL equations.  Thus we must ascribe a temperature 
relation to $s(t)$ so that our model elucidates the arguments for an increase 
or decrease of the critical temperature.  Since $s \ge 0$ by definition, the 
temperature function related to the E, F, and H solution types must be 
different from that of the A, C, and I solution types.  We introduce the 
relation 

\begin{equation}\label{tmodel}
s_{\pm}(t) = \sqrt{\pm\left({t^2-1\over t^2+1}\right)},~~~~~t = {T\over T_c},
\end{equation}
where the upper (lower) sign denotes the temperature model with 
$t > 1 ~(t < 1)$, respectively, and apply it to the exact results shown in 
Figs. \ref{fig2} and \ref{fig3}.  When $\chi(0) = 0$, the fundamental length 
ratios $s_{\pm}(t)$ assume the values $s_+(t_{cs})$ for $t_{cs} > 1$, while  
$s_-(t_{cs})$ for $t_{cs} < 1$.

At the reference temperature $t = 1$, the fundamental ratio $s(1) = 0$. For 
this model there are two temperature ranges $0 < t < 1$ and $1 < t < t_{cs}$, 
corresponding to the same value of $s$ in the interval $1 \ge s \ge 0$.  For 
$t_{cs} < 1$, an increasing $s(t)$ value corresponds to an increasing $\chi(0)$ 
value such as seen in the lower branches A and I in Figs. \ref{fig2} and 
\ref{fig3}.  The explicit equation for $s_-(t_{cs})$ is (\ref{bcAlin}), and 
fixed parameters $d/l_{\beta}$ and $d/b_2$, from which $t_{cs} < 1$ is obtained 
using Eq.\ (\ref{tmodel}).  For $t_{cs} >1$, an increasing $\chi(0)$ value 
corresponds to a decreasing $s(t)$ value such as is depicted in Fig. \ref{fig2} 
by the E branch and in Fig \ref{fig3} by the F, H1, and H2 branches.  In such 
cases the explicit equation for  $s_+(t_{cs})$ is Eq.\ (\ref{bcElin}), from 
which $t_{cs} > 1$ is obtained using Eq.\ (\ref{tmodel}).

With these $t_{cs}$ values, which are embedded in Eqs.\ (\ref{bcAlin}) and 
(\ref{bcElin}), the results of Figs. \ref{fig4} and \ref{fig5} are calculated 
from the $s(\chi(0))$ values shown in Figs. \ref{fig2} and \ref{fig3} using 
Eq.\ (\ref{tmodel}).  For the E, C2, C1, A curve in Fig. \ref{fig4} and the 
F, H1, H2, I curve in Fig. \ref{fig5}, the $\chi(0)$ curve is continuous and 
finite at $t = 1$, while $N_0 = f^2(0)$ is infinite at this point.  Again one 
sees that it is not the $f(x)$ function which is the correct physical order 
parameter but the $\chi(x)$ function. 

The A$|$C1 boundary point in Fig. \ref{fig4} and the F$|$H1 boundary point in 
Fig. \ref{fig5} are the uniform solution points, with corresponding temperature 
$t_u$.  Temperature $t_u$ may be found analytically using Eq.\ (\ref{sflat})
 with $k_j = 0$, and Eq.\ (\ref{tmodel}). The result is
\begin{equation}\label{tu}
t_u(\pm) =  \sqrt{{1 \mp  b_4/b_2\over 1 \pm b_4/b_2 }},
\end{equation}
where the upper\,(lower) sign denotes $t_u > 1 ~(t_u < 1)$, respectively.  
Using Eq.\ (\ref{tu}), the temperature $t_u = \sqrt{2/3}$ in Fig. \ref{fig4}, 
and $t_u = \sqrt{13/12}$ in Fig. \ref{fig5}.  

The conservative and convincing results shown by $\chi(0,t)$ lend credence to 
our results that boundary conditions imposed on small superconducting specimens 
may change drastically their behavior, including increasing the transition 
temperature. Changing the slope of the order parameter at the surface of thin 
films in a magnetic field parallel to the surface effects also the magnetic 
field at which superconductivity nucleates \cite{fink2}.

Although the temperature model, Eq.\ (\ref{tmodel}), is not unique, it appears 
to be in accord with experimental results described by curve A (lower curve) 
in Fig.\ref{fig4} when the transition temperature of a thin superconducting 
film is depressed by bringing it in contact with normal metals (NSN).  The 
upper curve in Fig. \ref{fig4} is also a reasonable result for a weaker 
superconductor S' in contact with a stronger superconductor S.  The model 
temperature behavior is different for the N-metal of a SNS junction: 
$s \propto \sqrt{T}$ (See Section IV).

%-----------------------------------------------------------------------------
\subsection*{Current Supporting Slab Solutions}

Next we consider the non-zero current solutions for $a  = 1$, which evolve 
continuously from the solutions plotted in Figs. \ref{fig2} and \ref{fig3} as 
the current parameter $k_j \propto j$ increases from zero.  Using the solutions 
from Eq.\ (\ref{solJ}) in the boundary condition (\ref{bc1DN}), with $a = 1$, 
yields the $\chi(0)$ versus $s(t)$ plots shown in Figs. \ref{fig6}-\ref{fig10}. 
 A typical nonlinear boundary equation that we solved is that for cases A, F, 
 and G. It is
\begin{equation}\label{bcJ} 
{1\over 2}{d\over l_\beta}{b_0\over\sqrt{\Delta}}{1\over N_s}
{\mbox{sn}(u_s)\mbox{cn}(u_s)\over \mbox{dn}^3(u_s)} + {d\over b_4}s^2 N_s = 
-{d\over b_2},
\end{equation}
where 
\[N_s = N(x = 0.5d) = N_0 + {b_0\over 2\Delta}\mbox{sd}^2(u_s),~~~~~~~u_s = 
{1\over 2}{d\over l_\beta}s\sqrt{\Delta},~~~~~~~m = 
{1\over 2}\left(1 + {\gamma \over\Delta}\right).\]

Figures \ref{fig6} and \ref{fig7} show the connectivity of all of the current 
carrying states A-E, with sgn$(\beta) = 1$, for various values of the unitless 
current parameter $k_j$. The bars denote the boundaries separating solution 
types.  The parameter set $(d/l_{\beta}, d/b_2, d/b_4)$, and the $k_j = 0$ 
curves are those depicted in Fig. \ref{fig2}.  In Fig. \ref{fig6} the 
temperature $t > 1$ is decreasing to $t = 1$ as $s$ decreases to $s = 0$.  
It is clear that a second order phase transition, characterized by $\chi = 0$, 
does not occur when $k_j \ne 0$. The upper curves in Fig. \ref{fig6} show the 
physical order parameter decreasing for a fixed $k_j$ as $s$ and the 
corresponding temperature increase; whereas the lower part of the curves 
behave unphysically.  At a fixed value of $s(t)$, increasing the current 
parameter $k_j$ decreases the order parameter\,(starting from the $k_j = 0$ 
curve) until a critical, maximum current is reached, above which the order 
parameter does not exist. This critical point is given by an E(D$_+$) solution 
for higher\,(lower) temperatures.  In Fig. \ref{fig7} the temperature 
$t < 1$ is decreasing from $t = 1$ as $s$ increases from $s = 0$.  Again, for 
a fixed $s(t)$ there is a critical, maximum current parameter $k_j$ above 
which the order parameter does not exist. The critical point is given by a 
D$_-$(A) solution for higher(lower) temperatures.  In Fig. \ref{fig7}, the 
horizontal line $\chi_u = \sqrt{1/5}$ is the value of the uniform solution, 
Eq.\ (\ref{sflat}).  The uniform solution is located at the A$|$C (A$|$B) 
boundary for higher (lower) temperatures.

Figure \ref{fig8} shows the A solution, which has a second order phase 
transition for $k_j = 0$, for various values of $k_j$. As $s$ increases from 
$s(t_{cs})$ the temperature $t< 1$ decreases. Other observations are analogous 
to those for Fig. \ref{fig7}.  

Figures \ref{fig9} and \ref{fig10} show the connectivity of all of the current 
carrying states F, G, H, and I, with sgn$(\beta) = -1$, for various values of 
$k_j$.  The parameter set $(d/l_{\beta}, d/b_2, d/b_4)$, and the $k_j = 0$ 
curves are those depicted in Fig.  \ref{fig3}.   In Figs. \ref{fig9} and 
\ref{fig10}, the horizontal line is $\chi_u = 1/5$, Eq.\ (\ref{sflat}), which 
is located at the F$|$H and G$|$I boundaries. Other significant observations 
are analogous to those for Figs.  \ref{fig6} and \ref{fig7}.
%----------------------------------------------------------------------------
\subsection*{Simple Analytic Solutions}
Here we return to the first order differential equation (\ref{dNdx}) to 
develop a simple algebraic model that very accurately fits the $k_j = 0$ curves 
for the cases A, F, H, and I, which have two finite extrema. Denoting the 
second extrema by $N_1 \equiv N(x_1)$, where $dN(x_1)/dx = 0 $, the term in 
the bracket $\{...\}$ in Eq.\ (\ref{dNdx}) is zero.  Using $\chi^2 = s^2N$ and 
$k_j = s^3J$ one obtains the exact relation between the extrema 
$\chi_0 \equiv  \chi(x_0)$ and $\chi_1 \equiv  \chi(x_1)$. It is

\begin{equation}\label{x1x0j} 
[\chi_0^2 + \chi_1^2 +2\mbox{sgn}(\alpha)\mbox{sgn}(\beta)s^2]\chi_0^2 \chi_1^2 
+2k_j^2\mbox{sgn}(\beta) = 0.
\end{equation} 
Consider the $k_j = 0$ cases with $\mbox{sgn}(\alpha)\mbox{sgn}(\beta) = -1$. 
Equation  (\ref{x1x0j}) reduces to

\begin{equation}\label{x1x0} 
\chi_0^2(s) + \chi_1^2(s) = 2s^2,~~~~\mbox{for}~~~\chi_0^2 \chi_1^2 \ne 0. 
\end{equation} 
It is evident that as $s \rightarrow 0$, the second extremum $\chi_1^2 = 
-\chi_0^2$.  Since $\chi_0^2$ is proportional to the physical electron pair 
density and is thus positive, $\chi_1^2(s=0) < 0$ is purely mathematical and 
does not represent a SC state.  Assume $\chi_1^2$ may be expanded in a power 
series in $s^2$.  Taking the leading terms, we write
\begin{equation}\label{x1} 
\chi_1^2(s) = a_0 + a_1s^2, 
\end{equation}   
which requires that 
\begin{equation}\label{xo} 
\chi_0^2(s) = -a_0 + (2-a_1)s^2. 
\end{equation}
The expansion coefficients are obtained by evaluating $\chi_0^2(s)$ at the 
SOPT point $s = s_o$, given by Eq.\ (\ref{bcAlin}) or (\ref{bcElin}), and at 
the uniform solution $\chi_0 = s_u = \sqrt{-b_4/b_2}$, which gives
\begin{equation}\label{ao} 
a_0 = -s_u^2\left[1-\left({s_u\over s_o}\right)^2\right]^{-1},~~~~a_1 = 
1 + \left[1- \left({s_u\over s_o}\right)^2\right]^{-1}.  
\end{equation}
When $s_u < s_o$, the coefficient $a_0 < 0$ and $a_1 > 2$.  This is the F case, 
and Eq.\ (\ref{xo}) lies within plot accuracy for all $s$ of the exact F-H1-H2 
curve shown in Fig. \ref{fig3}. When $s_u > s_o$, the coefficient $a_0 > 0$ 
and $a_1 < 2$. This is the A case, and  Eq.\ (\ref{xo}) lies within plot 
accuracy for all $s$ of the exact A-C1 SOPT curve shown in Fig. \ref{fig2}. 
Other curves that do not have a SOPT may also be obtained using 
Eq.\ (\ref{xo}); however the coefficients cannot be analytically evaluated. 

For $k_j \ne 0$ we have not found a simple expression for $\chi_1(s,k_j)$ such 
that Eq.\ (\ref{x1x0j}) produces the results shown in 
Figs. \ref{fig8}-\ref{fig10}.  It is evident from Fig. \ref{fig10}, for 
example, that an infinitesimal current completely changes the $\chi(0, s)$ 
curve.

%-----------------------------------------------------------------------------
\subsection*{Currentless Surface ``Pre-Wetting'' Solutions}
In this subsection we apply the solutions from Eq.\ (\ref{wetsol}), which exist 
only over a section of the slab.   Substituting the solutions of 
Eq.\ (\ref{wetsol}) into the boundary condition (\ref{bc1D}), with $a = 1$ 
gives the equations to be solved for the integration constant 
$C = 2(l_\alpha df/dx|x_w)^2$ for a given value of the surface pre-wetting 
depth $l_0 = 0.5d - x_w$.  A typical equation is that for the E' case:
 
\begin{equation}\label{bcwet} 
{d\over l_\beta}s\sqrt{{1+\sqrt{1-C}\over 2}}
{\mbox{dn}(u_s)\over \mbox{cn}(u_s)\mbox{sn}(u_s)} + 
{d\over b_4}s^2(1-\sqrt{1-C})\mbox{sc}^2(u_s) = -{d\over b_2},
\end{equation}
where
\[u_s = {l_0\over l_\beta}s\sqrt{{1+\sqrt{1-C}\over 2}},~~~~~m = 
{2\sqrt{1-C}\over 1+\sqrt{1-C}}.\]
If the $b_4$ surface interaction term is neglected, cases A' and E' are 
elliptic function representations of the extensive numerical solutions of 
Eq.\ (\ref{dfdxC}), with $J = 0$, evaluated in Ref. \onlinecite{montevecchi} 
by Indekeu and coworkers. 

For all cases, when the argument $u(x) \ll 1$, the function 
$f(x) \propto u(x)$.  In this limit, i.e. when the pre-wetting depth 
$l_0 \ll l_{\alpha}$, the boundary condition (\ref{bc1D}) leads to 
$l_0 = -b_2$. 

Cases A', E', F', and I' exhibit second order phase transitions(SOPT) in the 
limit $C \rightarrow 0$.  Then the boundary condition for A' and I' reduces 
to the form
 
\begin{equation}\label{soptwet1} 
{d\over l_\beta}s_o(t)\mbox{cot}\left[{l_0s_o(t)\over l_\beta}\right] = 
-{d\over b_2},~~~\mbox{for}~~f(x_w) = 0,
\end{equation} 
and for cases E' and F' it gives
\begin{equation}\label{soptwet2} 
{d\over l_\beta}s_o(t)\mbox{coth}\left[{l_0s_o(t)\over l_\beta}\right] = 
-{d\over b_2},~~~\mbox{for}~~ f(x_w) = 0.
\end{equation}
It is interesting to compare the pre-wetting solution characteristic 
equations for 
$l_0 = 0.5d$ with SOPT Eqs.\ (\ref{bcAlin}) and (\ref{bcElin}) for the slab 
solution cases A, I and E, F.  First note that in the A', I' cases $b_2 < 0$; 
whereas $b_2 > 0$ for the A, I cases. At the reference temperature, where
 $s(t = 1)\rightarrow 0$, Eq.\ (\ref{bcAlin}) for cases A, I requires 
 $1/b_2 \rightarrow 0$; whereas for A', I', Eq.\ (\ref{soptwet1}) gives 
 $|b_2| = l_0$ at $t = 1$.  Similar comments apply to cases E, F versus E', F'.

The pre-wetting solution normalized, minimum free energy, corresponding to 
Eq.\ (\ref{Gamin}), is

\begin{equation}\label{Gaminwet}
\Gamma_{min}(s) =  -{1\over 2}\left({dl_0\over l_{\beta}^2}\right)
{2\over l_0}\int^{0.5d}_{0.5d-l_0}dx\chi^4(x)\mbox{sgn}(\beta) - 
{d\over b_4}\chi^4(0.5d).
\end{equation} 

As an application, consider a slab of normal, or SC, material between two 
superconducting reservoirs with a $T_c$ higher than that of the slab. In the 
absence of a transport current, the order parameter is characterized by an 
A type solution in the interior of the reservoir with extremum value 
$f_0 \le 1$.  At the surface there will be some bending such that 
$f(d/2) <  f_0$.  For the slab we assume an E' type pre-wetting solution with 
surface value

\begin{equation}\label{fwet}
f^2(d/2) = \left( 1 - \sqrt{1-C}\right)\mbox{sc}^2(u_s|m)
\end{equation} 

Using Eq.\ (\ref{fwet}) with elliptic function identities to write the 
elliptic functions in the derivative boundary condition (\ref{bcwet}) in 
terms of $C$ leads to the expression
    
\begin{equation}\label{Cwet}
C(s) =f^2(d/2)\left[ 2\left({l_{\beta}\over b_2}{1\over s} + 
{l_{\beta}\over b_4}sf^2(d/2)\right)^2 -2-f^2(d/2)\right].
\end{equation} 
The value of $f(d/2)$ is a function of $s(t)$, but if one fixes the values of 
$f(d/2)$ in Eq.\ (\ref{Cwet}), the resulting $C(s)$ may be inserted in 
 Eq.\ (\ref{fwet}), to determine the pre-wetting depth $l_0$ as a function of 
 $s(t)$.

Fig. \ref{fig11} shows the pre-wetting depth $l_0$ plotted as a function of 
temperature, using $t^2 = (1-s^2)/(1+s^2)$ for several values of $f(d/2)$.  Near $T_c$ of the SC reservoirs the surface value $f(d/2) \ll 1$ and $l_0/d$ approaches $|b_2|/d = 0.2$.  As the temperature is decreased both $f(d/2)$ and $l_0$ have the tendency to increase until $l_0/d = 0.5$, at which point we expect a transition to the E type.

%==========================================================================
\section{Transport Currents in Multiple Layered Systems}

In the previous section we applied a boundary condition that was formulated 
from the surface energy characterized by the parameters $b_2$ and $b_4$, which 
were assumed to represent not only the surface properties, but also the 
material beyond the surface.  Now we formulate a set of boundary conditions 
that relate parameters across a boundary, bypassing the intermediate surface 
parameters. These boundary conditions are then applied to a SNS system with 
transport currents, and the results are compared with recent experiments.

%---------------------------------------------------------------------------
\subsection*{Crossing Boundaries}
There are three basic quantities that are relevant at a boundary: the surface 
energy in $G_{min}$, Eq.\ (\ref{Gmin1}), the boundary condition (\ref{gbc1}) 
with a = 1, and the current density ${\bf j}$, Eq.\ (\ref{gl2}).  The surface 
energy density is proportional to the function  
\begin{equation}\label{gs}
g_s =  -{1\over 2m^*}|\psi|^3\frac{\partial}{\partial |\psi|}
\left[{\Lambda\over |\psi|^2}\right] = {1\over m^*}\left[\Lambda - 
{1\over 2}|\psi| \frac{\partial\Lambda }{\partial |\psi|}\right],   
\end{equation}   
and the boundary condition (\ref{gbc1}) may be written in the form
\begin{equation}\label{bcs}
{1\over m^*}\left[\hat{\bf n}\cdot\nabla |\psi|^2 + 2\Lambda \right] = 2g_s.
\end{equation}   
Since $g_s$ is the same when viewed from either side of the surface at the 
same point on the surface, with $m^*$ changing value, in general, it follows 
that the right side of Eq.\ (\ref{bcs}) must be continuous across the surface. 
Assuming that there is no generation or loss of particles at the surface, we 
conjecture that at the boundary between medium 1 and medium 2, the function 
$\Lambda/m^*$ satisfies 
\begin{equation}\label{Lambc}
{\Lambda(1)\over m^*(1)}  = {\Lambda(2)\over m^*(2)}.
\end{equation} 
Continuity of $g_s$ and $\Lambda/m^*$ gives the continuity of the SC density 
gradient condition
\begin{equation}\label{gradbc}
{n_s^*(1)\over m^*(1)}\nabla \chi^2(1)|_b  = 
{n_s^*(2)\over m^*(2)}\nabla \chi^2(2)|_b.
\end{equation}

The second boundary condition is the continuity of the physical current 
density $j$. Using the definition (\ref{kj}),  continuity of $j$ requires that
\begin{equation}\label{kjbc}
 { n_s^*(1)\over m^*(1)l_{\beta}(1)} {\bf k}_j(1) = 
 { n_s^*(2)\over m^*(2)l_{\beta}(2)} {\bf k}_j(2).
\end{equation}

Aside from known material and geometry parameters, the functions on each side 
of Eq.\ (\ref{gradbc}) depend only on $s(t), k_j$ and $\chi(x_0)$, where 
$x_0$ is the reference point in the medium.  Thus, if $j$ and $t$ are fixed 
and the model $s_{\pm}$ is chosen for each medium, Eqs.\ (\ref{gradbc}) and 
(\ref{kjbc}) determine $\chi(x_0(1))$ in one medium as a function of 
$\chi(x_0(2))$.

In order to determine the value of the pair \{$\chi_0(1), \chi_0(2)$\}, 
i.e. to set the SC level, a third constraint is required. If the system is 
open, the boundary condition (\ref{bc1}) may applied at an outer surface; 
thus determining $\chi_0$ for the adjacent section as a function of the outer 
surface parameters.  For a closed system, such as a ring constructed from two 
materials, the second constraint may be the flux quantization condition, which 
follows from Eq.\ (\ref{Q}), using Eqs.\ (\ref{j}) and (\ref{kj}).  It is
\begin{equation}\label{fluxquant}
-\oint d{\bf l}\cdot\left({{\bf k}_j\over l_{\beta}\chi^2}\right) = 
2\pi \left(n + {1\over \phi_o}\phi(k_j)\right),
\end{equation}
where $\phi$ is the flux enclosed by the contour $l$.  
Equations (\ref{gradbc}) - (\ref{fluxquant}) determine the numerical value of 
$\chi(x_0)$ for each medium.

In lieu of imposing an outer boundary derivative condition in terms of 
presently unknown characteristic lengths $b_2, b_4$, one may set the ``level'' 
by simulating the outer surface with the imposition of a discontinuity in the 
physical pair density $|\psi|^2$ at the inner boundary.  That is, we set

\begin{equation}\label{psijump}
n_s^*(1)\chi^2(1) =\eta ~n_s^*(2)\chi^2(2),
\end{equation} 
where the parameter $\eta$ is a measure of the size of the jump in $|\psi|^2$. 
Equations (\ref{gradbc}), (\ref{kjbc}), and (\ref{psijump}) completely, and 
self-consistently determine the order parameter in each region of the system 
at a given value of $k_j$ and  $s(t)$ if the transition temperature of the 
system is known.

Finally we note that the continuity of the current density $j$, also relates 
the gradients of the phases of $\psi(1)$ and $\psi(2)$ via
\begin{equation}\label{Qbc}
{n_s^*(1)\over m^*(1)}\chi^2(1){\bf Q}(1)  = 
{ n_s^*(2)\over m^*(2)}\chi^2(2){\bf Q}(2).
\end{equation}

%------------------------------------------------------------------------
\subsection*{A Periodic SNS Layered System}

Here we analyze the experiments on SNS systems, AlAgAl and NbCuNb, as 
investigated by the Grenoble group \cite{dubos,courtois,dubos1}.   Although 
both the (F, H) and the (E, D$_+$) solution pairs are possible candidates for 
the N slab\,(Compare Figs. \ref{fig6} and \ref{fig9}), we choose the (E, D$_+$) 
solutions because they are pull up solutions throughout the entire temperature 
range, and because the positive value of $\beta$ is expected to give a lower 
energy state.  We believe that this choice is more likely to be the correct 
physical solution than that previously used\cite{fink4}, which corresponds to 
the F solution.  For simplicity, typical SC parameters and dimensions are used, 
similar to those used by Courtois et al.\cite{courtois} and 
Dubos et al.\cite{dubos,dubos1}

We denote the $superconducting$ layer (Al, Nb) properties by subscript ``$s$''
 and those of the $normal$ layer (Ag, Cu) by subscript ``$n$'' and use the 
 $N_0$ notation of Eq.\ (\ref{roottable}) for the extremum of the $f^2$ 
 function in the $s$ and $n$ layer.  Note that for case A, the restriction is 
 $0 \leq N_0 \leq 1$, while for $E$ and $D_{+}$ it is $0 \leq N_0 \leq \infty$.

We assume that the current in the $n$ layer $k_{jn}  = 
 - s_n^3(t)J_n$ is phase-coherent and is treated as a pair current as it is in 
 the superconducting
layers. We define in the $n$ layer "coherence'' length  
$\xi _n(t) = l_{\alpha n} = l _{\beta n} /s_n$, with 
$l _{\beta n}^2  = (\hbar/2\pi )(D/k_B T_c)$, where 
$3D = v_{Fn}\ell _n  = \ell _n^2 /\tau$, with $D$ the diffusion constant, 
$\ell_n$ the mean free path, $v_{Fn}$ the Fermi velocity, and $\tau$ the 
electron scattering time.  The transition temperature of the SNS structure is 
$T_c$, $k_B$ Boltzmann's constant and $s_n  = \sqrt {t} = \sqrt {T/T_c }$.

In the $s$ layer the coherence length is $\xi _s(t) = l_{\alpha s} = 
l _{\beta s}(t)/s_s  = \xi _s (0)(1 + t^2)/s_s$, with $s_s  = \sqrt{1 - t^4}$. 
 The effective  penetration depth $\lambda _{eff}  = 
 \Omega \lambda _L /\sqrt{1 - t^4}$, where $\lambda_{L}$ is the London 
 penetration depth and $\Omega  = 0.65(\xi _0 /\lambda _L )^{1/3}  > 1$ for Al, 
 a Pippard superconductor, for which $\xi _s^3 (0) \gg \xi _0 \lambda _L^2$ 
 ($\xi_0 = $ BCS coherence length) is satisfied.  For $Nb$  in the dirty limit 
 $\Omega  = (\xi _0 /\xi _s (0))^{1/2}  > 1$, with
 
\[\xi _s(0) = \frac{\phi_o}{2\pi\sqrt{2}\mu _0 H_c(0)\lambda _{eff}(0)},\]
where $\phi_o$ is the SC fluxoid quantum, $H_c(t)  = H_c(0)(1 - t^2)$ is the 
thermodynamic critical field and $\mu_0$ the vacuum 
permeability.  The Drude resistance of the $n$ layer $R_n  = 
[m_n /(n_n e^2\tau)](d_n /A)$, with $d_n$ the thickness of the normal layer, 
$A$ the cross-sectional area.

The physical current density $j$ is  the same in the $n$ and $s$ regions, and 
thus the relation between $j$ and the normalized current densities in the $n$ 
and $s$ regions is
\begin{equation}\label{dJbc}
J_n  = \left(\frac{s_s}{s_n} \right)^3{1\over r}\frac{l_{\beta n}}
{l_{\beta s}}J_s  = \left(\frac{s_s}{s_n}\right)^3 {1\over r}\frac{l_{\beta n}}
{\xi _s(0)(1 + t^2)}\frac{\lambda _L}{\sqrt{2}H_c(t)\,\Omega\,s_s }j,
\end{equation}
where $r = (m_s /m_n)(n_n /n_s)$.  From the above, the physical critical 
current density $j_c$, Eq.\ (\ref{kj}), leads to the energy expression
\begin{equation}\label{ej}
 ej_cA\,R_n  = {1\over 2}\frac{\hbar}{\tau}\left(\frac{d_n}
{l _{\beta n}} \right)k_{jc} = \varepsilon {3\over 2}\frac{d_n}
{l _{\beta n}}\left({d_n\over \ell_n}\right)^2 k_{jc}
\end{equation}
where $\varepsilon  = \hbar\,D/d_n^2$ is the Thouless energy.

It is our aim to calculate the largest (critical) current density 
$k_{jc}(t) = \mbox{max}[k_{j}]$ of a SNS structure as a function of 
temperature.  The prefactor of $k_{jc}$ depends solely on properties of the 
$n$ layer, but the unitless current parameter $k_{jc}$ tracks the physical 
critical current density $j_c$ and is implicitly connected to the SC and 
normal region parameters.   The $E$ and $D_{+}$ solutions applicable to the 
$n$ region, and $A$ solution relevant to the $s$ region are, using 
Eq.\ (\ref{solJ})

\begin{equation}\label{solsns}
\begin{array}{ll}

\mbox{A} &  N_s(x) = N_0 + \displaystyle{{1\over 2}
{b_0\over \Delta}}\mbox{sd}^2(u|m)\\[.3cm] 

\mbox{E} & N_n(x) = N_0 + 
(1 +\displaystyle{{3\over 2}}N_0 - \Delta)\mbox{sc}^2(u|m)\\[.3cm]

\mbox{D}_+ &  N_n(x) = N_0 +  \sqrt{b_0}~\mbox{sc}^2(u|m)\mbox{dn}^2(u|m).
\end{array}
\end{equation}

 For simplicity, we set $m_s /m_n  = n_s /n_n = 1$ which reduce
  Eqs.\ (\ref{gradbc}) and (\ref{psijump}), respectively, to 

\begin{equation}\label{dbcN}
\frac{dN_n}{dx}|_b  = \left[{s_s(t)\over s_n(t)} \right]^2 \frac{dN_s}{dx}|_b 
\end{equation}
and
 \begin{equation}\label{bcN}
N_n(b) = \eta \left[{s_s(t)\over s_n(t)}\right]^2 N_s(b).
\end{equation}
Although the parameter $\eta$ is set by ``external'' boundary conditions, 
we assume continuity of the physical pair density across the boundary, i.e. 
$|\psi_s|_b^2  = |\psi_n|_b^2$, with $\eta = 1$.  

Since $u$ and $m$ are functions of $N_0$ and $J$, it is possible to calculate 
for any pair $(N_0, J)$ the values of $N(x = d/2) = N_b$ and $dN/dx|(x = d/2)$ 
at the interface between the $n$ and $s$ regions. The triple $(N_0, J, N_b)$ 
comprise a surface, and it is then possible to plot $J$ vs. $N_0$ for fixed 
$N_b$.

The $D_+$ solutions for several values of $N_b$ for a NbCuNb specimen at 
$T = 0.35\,K$ are shown in Fig. \ref{fig12}.  The $D_{+}$ solutions terminate 
on the right and the E solutions continue to $J = 0$ [not shown].  For fixed 
$N_0$ and $N_b$ there exists a maximum $J$ which is proportional to the 
critical current density $j_c(t) \propto k_{jc}(t)$, provided boundary 
conditions (\ref{kjbc}), (\ref{dbcN}), and (\ref{bcN}) are satisfied.  Only 
the curve with $N_b = 16.73$ satisfies all three boundary conditions at 
$T = 0.35K$; thus the maximum value of this curve, which occurs in the $D_+$ 
solution region at $T = 0.35K$, corresponds to the unique $j_c$.   Similarly, 
Fig. \ref{fig13} shows E solutions at $T = 0.5\,K$ for the above NbCuNb 
specimen. The E solutions terminate on the left and continue as $D_{+}$ 
solutions [not shown].  Only the $N_b = 11.63$ curve satisfies all boundary 
conditions, and therefore its maximum value, which lies in the E solution 
region, is the correct unique solution for the critical current at 
$T = 0.5\,K$.  The general behavior depicted in Figs. \ref{fig12} and 
\ref{fig13} is consistent with results expected from Fig. \ref{fig6} for 
constant temperature.

Critical current values $k_{jc} \propto j_c$, Eq.\ (\ref{kj}), are plotted in 
Fig. \ref{fig14} as a function of normalized temperature for the above NbCuNb 
and AlAgAl junctions with parameters listed in the captions for 
Figs. \ref{fig12} and \ref{fig14}.  The first three points on the left of each 
curve are D$_+$ solutions, and the remaining points are E solutions. The 
plotted points are exact solutions with all three boundary conditions 
satisfied.  The solid line is an aid to the eye only, connecting the computed 
solutions.  The extrapolated values of $k_{jc}$ to $T = 0~$K of the two 
junctions are $8.7\times 10^{-3}$ and $6.5\times 10^{-3}$, and are comparable 
to the maximum $k_j$ value $\sim 7\times 10^{-3}$ in Fig. \ref{fig6}, although 
they were obtained with different external parameters.  Similarly, the maximum 
value of $k_j$ for the F type solution shown, in Fig. \ref{fig9}, is 
$8\times 10^{-3}$. These results add credence to our earlier statements that 
it is $k_{j} \propto j$, and not $J$, that is a meaningful parameter.  

In Ref. \onlinecite{dubos1} the dimensionless parameter $eI_cR_n/\varepsilon$, 
Eq.\ (\ref{ej}), is plotted versus $k_BT/\varepsilon$, with $\varepsilon$ a 
fitting parameter. These experiments indicate that this parameter extrapolated 
at $T = 0~$K to the value $8.2$.  With the above parameters, the $0~$K 
limiting value of $eI_cR_n/\varepsilon$ for the NbCuNb junction is $47$, and 
for the AlAgAl junction $257$, indicating that this value depends strongly on 
the junction parameters of Eq.\ (\ref{ej}).

Figure \ref{fig15} shows the experimental points, taken from Fig. 4 of 
Ref. \onlinecite{dubos}, of critical currents of a NbCuNb junction with 
parameters similar to those of Fig. \ref{fig12}.  The solid line is calculated 
from the $k_{jc}(t)$ curve of Fig. \ref{fig14} using Eq.\ (\ref{kj}) with 
$e^* = 2|e|$, $m^*/n_s^* = 1.12\times 10^{-58}~$kg\,m$^3$, and 
$A = 5.9\times 10^{-14}$m$^2$.  The latter curve is a convincing fit of our 
theory to the experimental points.  Assuming that the extrapolation to 
$T = 0$\,K is meaningful, one obtains $eI_cR_n/\varepsilon = 21$, using 
$I_c(0) = 2.5$\,mA,  $\varepsilon = 24\,\mu$eV, and $R_n = 0.20\,\Omega$.

%=============================================
\section{Conclusions}
We have reformulated the GL theory by introducing the complete kinetic energy 
density, which requires a gradient term in the surface energy to support a SC 
state, thus replacing the standard GL energy density functional.
The role of surface energy in determining the superconducting state of a 
sample was analyzed in detail.  For a weakly superconducting surface, we have 
shown that the same phenomenological parameter $b_2$ can be used to describe a 
reduction or an enhancement of the transition temperature in moderately small 
to small superconductors. Plating a superconducting specimen with a normal 
metal or ferromagnetic substance reduces the transition temperature and is 
described by $b_2 > 0$.  For $b_2 < 0$ an increase of $T_c$ is perhaps brought 
about by elastic strain\cite{bibby}, observed on tin whiskers,  by severely 
cold working the surface of {\rm InBi} foils\cite{fink}, and by plating the 
specimen with a superconductor with a larger intrinsic $T_c$ than the specimen, 
and by other means\cite{astrak,fang,mishonov,buzdin}. In recent theoretical
 studies\cite{sigrist} the $3$ Kelvin phase of ${\rm Sr_2RuO_4}$ is modeled 
 with {\rm Ru} metal inclusion as interface states with locally enhanced 
 ($b_2 < 0$) transition temperature. A negative $b_2$ value, 
or more precisely, a positive slope of the order parameter imposed at the 
surface of a normal slab, embedded between superconductors, induces 
superconductivity throughout the normal region if sufficiently thin.  
Nucleation fields for slabs for positive and negative slopes of the order 
parameter at the slab surfaces have been published previously\cite{fink2}.  
They show that enhancements of $H_{c3}$ for $b_2 < 0$ is also possible. Other 
practical surface treatments and theoretical microscopic explanations relating 
to $T_c$ enhancements of superconductors  are still  to be discovered. 

The parameter $\beta$, characterizing SC pair interactions in a sample, is 
always positive when the surface is weakly superconducting.  If the surface is 
strongly superconducting, that is, when SC pair interactions play a significant 
role, superconductivity of the system is also characterized by a parameter 
$b_4$.   In this case, we have shown that the sample may be in an anomalous 
superconducting state even when $\beta$ is negative.  The parameter $b_2$ can 
be measured from SOPT experiments; whereas the parameter $b_4$ can be 
determined from the uniform state, if it can be detected, or perhaps from the 
minimum surface free energy. 

For one dimensional systems with a uniform current density, all possible 
physical solutions of our generalized GL equations with sgn$(\alpha ) = \pm 1$ 
and sgn$(\beta) = \pm 1$ were found and categorized.  Although a commonly used 
parameter $f$ is mathematically expedient, it can be misleading when used to 
depict the SC order parameter when surface effects are significant, i.e. $f$ is 
not, in general, a physical order parameter.

Introducing a transport current modifies the solutions of the theory with 
arbitrary parameters $\alpha$ and $\beta$.  Figures \ref{fig6} - \ref{fig10}, 
in which the order parameter $\chi =\psi /\sqrt{n_s^*}$ is plotted as a 
function of the fundamental length ratio $s = l_{\beta}/l_{\alpha}$, 
Eq.\ (\ref{s}), relate to solutions with currents which are described by 
Eq. (\ref{solJ0}), while Eq.\ (\ref{wetsol}) and Fig. \ref{fig11} are zero 
current pre-wetting solutions which evolve naturally in the present 
development.  The current parameter $k_j$, Eq.\ (\ref{kj}), tracks the 
physical current density.  Since $k_j$ remains finite as $s(t) \rightarrow 0$, 
the critical current remains finite in a normal region, where 
$s(t) = \sqrt{t}$, as $t \rightarrow 0$.  In Figs. \ref{fig12}-\ref{fig15} the 
transport current results, applied to SNS junctions, are in excellent 
agreement with experiments over the entire wide temperature range of the 
measurements.  The boundary condition for the gradient of the order parameter 
at the interface is derived from the surface energy functional. This boundary 
condition is consistent with that used in micronetworks\cite{degennes2}.  The 
continuity of the order parameter at the interface, assumed without formal 
justification, completes the set of boundary conditions necessary to set the 
``level'' of the order parameter.  Although this condition is consistent with 
that used in quantum mechanics, the level of the order parameter is set, in 
general, by an external boundary condition on its gradient.  

It is often argued that the GL model equations apply only near $T_c$.  However, 
 this argument is weak, since the GL phenomenological model does not contain 
 the temperature explicitly.  The exact solutions given here for the order 
 parameter are functions of the fundamental length ratio $s(t)$.  It is true, 
 if we write $s = \sqrt{1-t}$, as GL did, our results would be limited to 
 temperatures near $T_c$.  However, there is no compelling requirement that 
 restricts the temperature dependence of the present theory to this limited 
 temperature range. The Gor'kov derivation \cite{gorkov} of the GL equations 
 uses a small energy gap expansion of the microscopic BCS theory\cite{BCS}, 
 valid near $T_c$, but that does not exclude the possibility that the 
 phenomenological order parameter theory gives reasonable physical results 
 well below\cite{deo} $T_c$.  However, it is prudent to relate $s(t)$ 
 to the experimental results of the temperature dependence $l_{\alpha}(t)$ and 
 $l_{\beta}(t)$.
%------------------------------------------------------------------------
\begin{appendix}
\section{Minimization of  the functional $G$}

The energy functional Eq.\ (\ref{G1}) is

\begin{eqnarray}\label{GA}
G& = & \int_V d^3{\bf r}\left[U(|\psi|)  + 
{\hbar^2\over 2m^*}[\nabla|\psi|\cdot\nabla|\psi| + 
|\psi|^2{\bf Q}\cdot{\bf Q}] + {1\over 2\mu_o}(\nabla\times{\bf A}')^2\right] + 
\\
 & &{\hbar^2\over 2m^*}\int_S d{\bf s}\cdot[\hat{\bf n}\Lambda +
 (c- \epsilon )|\psi|\nabla|\psi|]\nonumber. 
\end{eqnarray}
where ${\bf A}' = {\bf A} - {\bf A}_a$.

The total variation of $G$ is
\begin{eqnarray}\label{varGA}
{2m^*\over \hbar^2}\delta G & = 
&2 \int_V d^3{\bf r}\{[{m^*\over \hbar^2}\frac{\partial U}{\partial |\psi|} + 
|\psi|{\bf Q}\cdot{\bf Q}]\delta |\psi| + 
\nabla|\psi| \cdot\nabla\delta |\psi| + |\psi|^2{\bf Q}\cdot\nabla\delta\theta 
+ \nonumber\\
 & &  \epsilon_s {2\pi\over \phi_o}|\psi|^2{\bf Q}\cdot\delta{\bf A} +
 {m^*\over \mu_o\hbar^2}(\nabla\times{\bf A}')
 \cdot(\nabla\times\delta{\bf A})\}  + \\
& & \int_S d{\bf s}\cdot[\hat{\bf n}\delta\Lambda +
(c-\epsilon)\delta(|\psi|\nabla|\psi|)]\nonumber.  
\end{eqnarray}
Consider the vector identities
\begin{eqnarray}
\nabla f\cdot\nabla g & = & - f\nabla\cdot\nabla g + 
\nabla\cdot(f\nabla g)\label{ident1}\\
{\bf B}\cdot(\nabla\times{\bf A}) & = & {\bf A}\cdot(\nabla\times{\bf A}) + 
\nabla\cdot[{\bf A}\times{\bf B}]\label{ident2}
\end{eqnarray}
Using the vector identities for the volume terms involving 
$\nabla\delta |\psi|, \nabla\delta \theta, \nabla\times\delta {\bf A}$, and 
applying the divergence theorem, one obtains from Eq.\ (\ref{varGA}) the 
variational form 
\begin{eqnarray}\label{varGA1}
{2m^*\over \hbar^2}\delta G & = 
& 2\int_V d^3{\bf r}\{[{m^*\over \hbar^2}\frac{\partial U}{\partial |\psi|} + 
|\psi|{\bf Q}\cdot{\bf Q}-\nabla^2 |\psi|]\delta |\psi| - 
\nabla\cdot(|\psi|^2{\bf Q})\delta\theta  + \nonumber\\ 
& &  [{2\pi\over \phi_o}|\psi|^2{\bf Q} + 
{m^*\over \mu_o\hbar^2}\nabla\times(\nabla\times{\bf A}')]\cdot\delta{\bf A}\}
 + \\
 & & \int d{\bf s}\cdot\{[(a + 1)\nabla|\psi|+
 (a-1) |\psi|\frac{\partial\nabla |\psi|}{\partial |\psi|} + 
 \hat{\bf n}{\partial \Lambda\over\partial |\psi|} ]\delta|\psi|  + 
 [ 2|\psi|^2{\bf Q} + 
 \hat{\bf n}{\partial \Lambda\over\partial\theta}]\delta\theta + \nonumber\\
 & & \hat{\bf n} [{2m^*\over \mu_o\hbar^2}(\nabla\times{\bf A}')
 \times \hat{\bf n} + \nabla_{\bf A}\Lambda]\cdot\delta{\bf A}\},\nonumber  
\end{eqnarray}
where $a = 1 + c - \epsilon$.  The variations $\delta |\psi|,\delta\theta, 
\delta{\bf A}$ may be taken independently in $V$ and on $S$. Thus the 
coefficients of the variations are zero and one obtains 
Eqs.\ (\ref{geq1})-(\ref{gbc3}).

To obtain the form of the minimum free energy given by Eq.\ (\ref{Gmin}), 
use (\ref{ident1}) with $g$ replaced by $f$ to eliminate 
$\nabla f\cdot\nabla f$, and use Eq.\ (\ref{geq1}) to eliminate 
$\nabla^2 |\psi|$. 

\end{appendix}
%=========================================================================

\begin{figure}
\caption{\label{fig1}The order parameter $\chi(0)$ of the E solution for is 
plotted as a function of the fundamental length ratio 
$s(t) = l_{\beta}/l_{\alpha}(t)$ for $a = 1$ and for $ a = 1 \pm 0.001$, 
where $a$ is a measure of the amount of $\nabla\chi^2$ in the surface energy. 
The dashed curves are the energy parameter $\Gamma_{min}$, Eq.\ (\ref{Gamin}). 
Only the $a = 1$ curve satisfies the entropy condition (\ref{sopt}) for a 
second order phase transition } 
\end{figure}

\begin{figure}
\caption{\label{fig2}The order parameter $\chi(0)$ is plotted as a function of 
$s$ for positive $\beta$ cases A, C1, C2, and E for three surface parameter 
sets $(b_2, b_4)$.  Also shown by dashed curves are the normalized, minimum 
free energy $\Gamma_{min}$.  A superconducting state requires 
$\Gamma_{min} < 0$.} 
\end{figure}

\begin{figure}
\caption{\label{fig3}The order parameter $\chi(0)$ is plotted as a function of 
$s$ for the negative $\beta$ cases F, H1, H2, and I for two surface parameter 
sets $(b_2, b_4)$. The dashed lines are the minimized energy $\Gamma_{min}$, 
Eq.\ (\ref{Gamin}).  The superconducting state requires $\Gamma_{min} < 0$.}
\end{figure}

\begin{figure}
\caption{\label{fig4}The order parameter $\chi(0)$ for the cases shown in 
Fig. \ref{fig2} are plotted as a function of the normalized temperature 
$t = T/T_c$ using Eq.\ (\ref{tmodel}).} 
\end{figure}

\begin{figure}
\caption{\label{fig5}The order parameter $\chi(0)$ for the cases shown in 
Fig. \ref{fig3} are plotted as a function of the normalized temperature 
$t = T/T_c$ using Eq.\ (\ref{tmodel}).} 
\end{figure}

\begin{figure}
\caption{\label{fig6}The order parameter $\chi(0)$ is plotted as a function of 
$s$ for positive $\beta$ cases E and D$_+$ for different values of the current 
parameter $k_j$. The $k_j = 0$ curves is that in Fig. \ref{fig2}, curve E.} 
\end{figure}

\begin{figure}
\caption{\label{fig7}The order parameter $\chi(0)$ is plotted as a function of 
$s$ for positive $\beta$ cases A, B, C, and D$_-$ for different values of the 
current parameter $k_j$. The $k_j = 0$ curve is that in Fig. \ref{fig2}, 
upper branch C2, C1, A.} 
\end{figure}

\begin{figure}
\caption{\label{fig8}The order parameter $\chi(0)$ is plotted as a function of 
$s$ for positive $\beta$ case A\,(lower branch) for different values of the 
current parameter $k_j$. The $k_j = 0$ curve is the lower branch in 
Fig. \ref{fig2}.} 
\end{figure}

\begin{figure}
\caption{\label{fig9}The order parameter $\chi(0)$ is plotted as a function of 
$s$ for positive $\beta$ cases F and H for different values of the current 
parameter $k_j$. The $k_j = 0$ curve is the F, H1, H2 curve in 
Fig. \ref{fig3}.} 
\end{figure}

\begin{figure}
\caption{\label{fig10}The order parameter $\chi(0)$ is plotted as a function of 
$s$ for positive $\beta$ cases G and I for different values of the current 
parameter $k_j$. The $k_j = 0$ curve is the upper I branch in Fig. \ref{fig3}.} 
\end{figure}

\begin{figure}
\caption{\label{fig11}Shown is the pre-wetting depth $l_0$ of the E' solution, 
 Eq.\ (\ref{wetsol}), plotted as a function of temperature for fixed values of
  $f(d/2)$ at the NS interface of a SC reservoir.}
\end{figure}

\begin{figure}
\caption{\label{fig12}
Normalized current $J$ vs. the minimum of the squared order parameter 
$f_0^2  = N_0$ in the $n$ region at constant temperature for various 
$N_b  = f_b^2$ values at the SN interface for the $D_+$ solution of NbCu at 
$T = 0.35~$K.  The continuation of these curves on the right are 
E solutions\,[not shown].  Parameters: $d_n = 800\,${nm}, $d_s = 200\,$nm, 
$l_{\beta n}= 61.3\,$nm, $l_{\beta s}(0) = 22.1\,$nm, $l_n = 48\,$nm, 
$\lambda_L = 39\,$nm, $\xi_{0}= 38\,$nm, $T_{c}=8.1\,$K, 
$v_{Fn} = 1.57\times10^{6}\,$m/s.}
\end{figure}

\begin{figure}
\caption{\label{fig13}
Similar to Fig. \ref{fig12}, but for the E solution of NbCu at $T= 0.5~$K.  
The continuation of these curves on the left are $D_+$ solutions\,[not shown].  The parameters are the same as used in Fig. \ref{fig12}.}
\end{figure}

\begin{figure}
\caption{\label{fig14}
Normalized critical current parameter $k_{jc} \propto j_c$, Eq.\ (\ref{kj}), 
plotted as a function of normalized temperature for a NbCuNb junction with 
the parameters of Fig. \ref{fig12}, and the same for an AlAgAl junction with 
parameters:  $d_n  = 1.48\,\mu$m, $d_s  = 2d_n,\;
l _{\beta n}  = 113\,$nm, $l_{\beta s}(0) = 487\,$nm, 
$l_{n} = 33\,$nm, $\lambda_{L}=14.8\,$nm, $\xi _0  = 1.6\,\mu$m, 
$T_c  = 1.45\,$K, $v_{Fn}  = 1.38 \times 10^6\,$ m/s.}
\end{figure}

\begin{figure}
\caption{\label{fig15}
Critical current $I_c$ plotted as a function of temperature $T$ of a NbCuNb 
junction.  The experimental points are from Ref. 7, and the solid line is 
calculated from $k_{jc}$ of Fig. \ref{fig14}.}
\end{figure}


\begin{references}

\bibitem{ginzburg} V.L. Ginzburg and L.D. Landau, 
Zh. Eksperim. i. Teor. Fiz. {\bf 20}, 1064 (1950)

\bibitem{haley} S.B. Haley, 
Phys. Rev. Lett. {\bf 74}, 3261 (1995); 
S.B. Haley and H.J. Fink,  Phys. Rev. B {\bf 53}, 3506 (1996)

\bibitem{degennes} P. G. de Gennes, 
{\it Superconductivity of Metals
 and Alloys}, W. A. Benjamin, New York (1966), 
reissued by Addison-Wesley, New York, (1992). 
[see also: {\rm Rev. Mod. Phys.} {\bf 36}, 225 (1964)].

\bibitem{simonin} J. Simonin,  Phys. Rev. B {\bf 33}, 7830 (1986).

\bibitem{indekeu} J. O. Indekeu and J.M.J. van Leeuwen, 
{\rm Phys. Rev. Lett.} {\bf 75}, 1618 (1995); 
{\it ibid.} {\rm Physica C} {\bf 251}, 290 (1995).

\bibitem{fink} H. J. Fink and W. C. H. Joiner,
 {\rm Phys. Rev. Lett.} {\bf 23}, 120 (1969).

\bibitem{dubos} P. Dubos, H. Courtois, O. Buisson, and B. Pannetier, 
 Phys. Rev. Lett. {\bf 87}, 206801 (2001).
 
\bibitem{fath} G. F\'{a}th and S.B. Haley,  Phys. Rev. B  {\bf 58}, 1405 (1998).
\bibitem{haley1} S.B. Haley, H.J. Fink, and  G. F\'{a}th, 
 Phys. Rev. B  {\bf 61}, 4353 (2000).

\bibitem{gorkov} L. P. Gor'kov, 
Zh. Eksperim. i Teor. Fiz. {\bf 36}, 1918 (1959); 
Soviet Phys. JETP {\bf 9}, 1364 (1959).

\bibitem{BCS} J. Bardeen, L.N. Cooper, and J. R. Schrieffer,  
Phys. Rev. {\bf 108}, 1175 (1957).

\bibitem{degennes2} P. G. de Gennes, 
C. R. Acad. SCi, Ser. II, {\bf 292}, 279, (1981); 
J. Simonin, D. Rodrigues, and A. L\'{o}pez, Phys. Rev. Lett. 
{\bf 49}, 944 (1982); S. Alexander, Phys. Rev. B {\bf 27}, 1541 (1983);
 H. J. Fink and S. B. Haley, Phys. Rev. Lett. {\bf 66}, 216 (1991);
  L. F. Chibotaru, A. Ceulemans, V. Bruyndoncx, and 
  V. V. Moshchalkov, Phys. Rev. Lett. {\bf 86}, 1323 (2001). 


\bibitem{tinkham} M. Tinkham, 
{\it Introduction to Superconductivity}, (McGraw-Hill, New York, 1996) 2nd Ed. 

\bibitem{ammann} C. Ammann, P. Erd\"{o}s, and S.B. Haley,  
Phys. Rev. B  {\bf 51}, 11739 (1995);  
S. Clark, P Erd\"{o}s, and H. J. Fink, 
{\rm Supercond. Sci. and Technol.} {\bf 13}, 1309 (2000).

\bibitem{montevecchi} E. Montevecchi, and J. O. Indekeu , 
Phys. Rev. B {\bf 62}, 14359 (2000); E. Montevecchi and J. O. Indekeu, 
{\rm Europhys. Lett.} {\bf 51}, 661 (2000); {\it ibid.}, 
{\rm Phys. Rev. B} {\bf 62}, 14\,359 (2000); F. Clarysse, 
Doctoral Dissertation (2000), Natuurkunde, Katholieke Universiteit, 
Louven, Belgium.

\bibitem{fink2} H. J. Fink, {\rm  J. Low Temp. Phys.} {\bf 16}, 387 (1974).

\bibitem{courtois}H. Courtois,P. Gandit, and B. Pannetier, 
{\rm Phys. Rev. B} {\bf 52}, 1162 (1995). 

\bibitem{dubos1} P. Dubos, H. Courtois, B Pannetier, F. K. Wilkhelm, A.D. Zaikin,
 and G. Sch\"{o}n,  {\rm Phys. Rev. B} {\bf 63}, 064502 (2001).

\bibitem{fink4} H. J. Fink, {\rm Phys. Rev. B} {\bf 56}, 2732 (1997). 

\bibitem{bibby} B. D. Rothberg--Bibby, H. J. Fink, D. S. McLachlan, 
and F. R. N. Nabarro, {\rm J. Low Temp. Phys.} {\bf 53}, 375 (1983).
                                           
\bibitem{astrak}   E. G. Astrakharchik and C. J. Adkins, 
{\rm J. Phys.: Cond. Matt.} {\bf 10}, 4509 (1998).  

\bibitem{fang} M. M. Fang, V. G. Kogan, D. K. Finnemore, J. R. Clem,
 L. S. Chumbley, and D. E. Farrell, 
 {\rm Phys. Rev. B} {\bf 37}, 2334 (1988); 
 L. A. Schwartzkopf, M. M. Fang, L. S. Chumbley, and D. K. Finnemore, 
 {\rm Physica C} {\bf 153-155}, 1463 (1988).

\bibitem{mishonov} N. B. Ivanov and T. M. Mishonov, 
{\rm Phys. Stat. Sol. (b)} {\bf 142}, K49 (1987).

\bibitem{buzdin} I. N. Khlyustikov and A. I. Buzdin, 
{\rm Advances in Physics} {\bf 36}, 271 (1987); F. Clarysse and J. O. Indekeu,
{\rm Phys. Rev. B} {\bf 65}, 094515, 11p. (2002); cond-mat/0106294.

\bibitem{sigrist} M. Sigrist and H. Monien,
{\rm J. Phys. Soc. Japan} {\bf 70}, 2409 (2001).

\bibitem{deo} P. S. Deo, V. A. Schweigert, F. M. Peeters, and A. K. Geim, 
{\rm Phys. Rev. Lett.} {\bf 79}, 4653 (1997). 



\end{references}
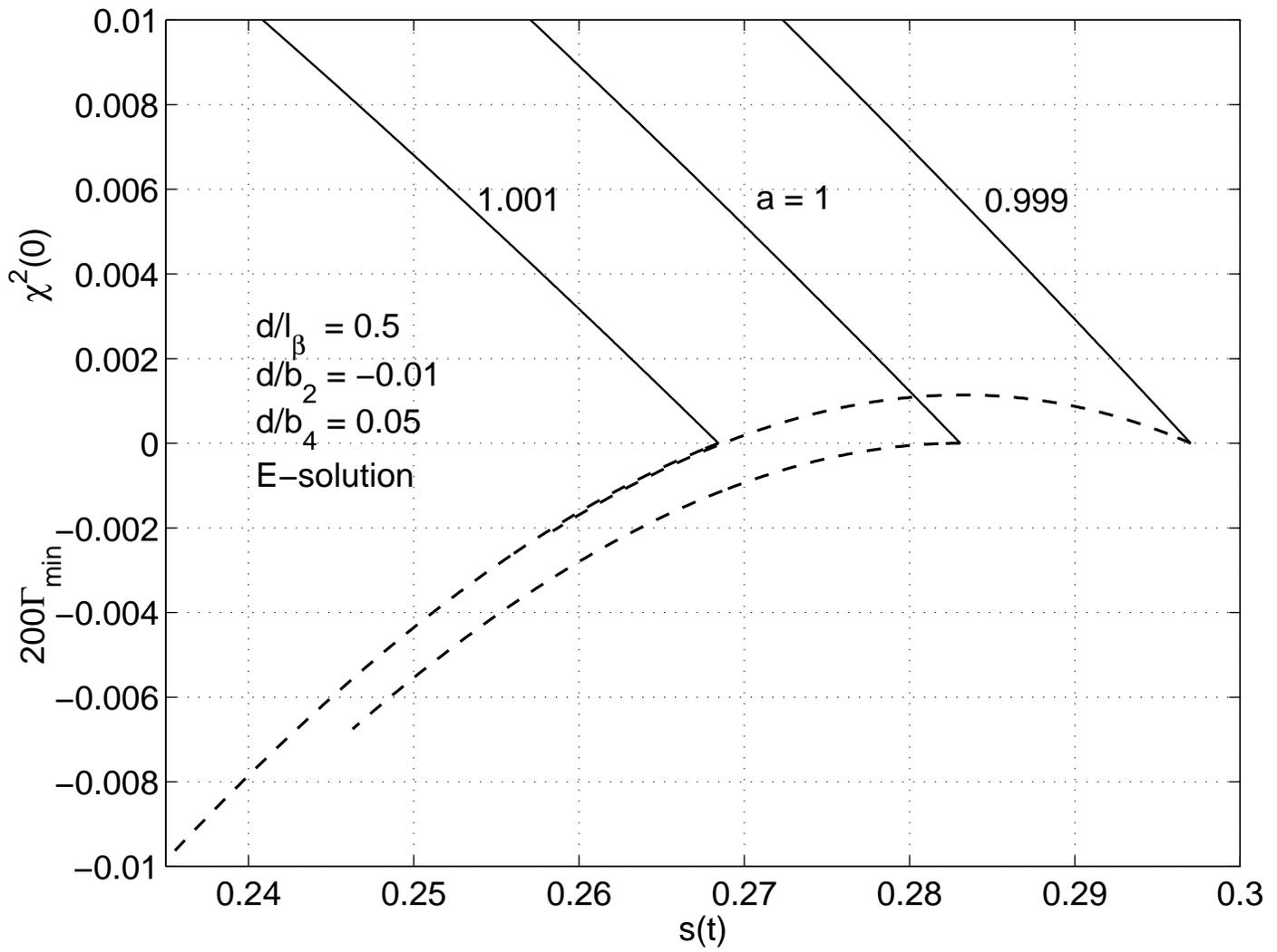
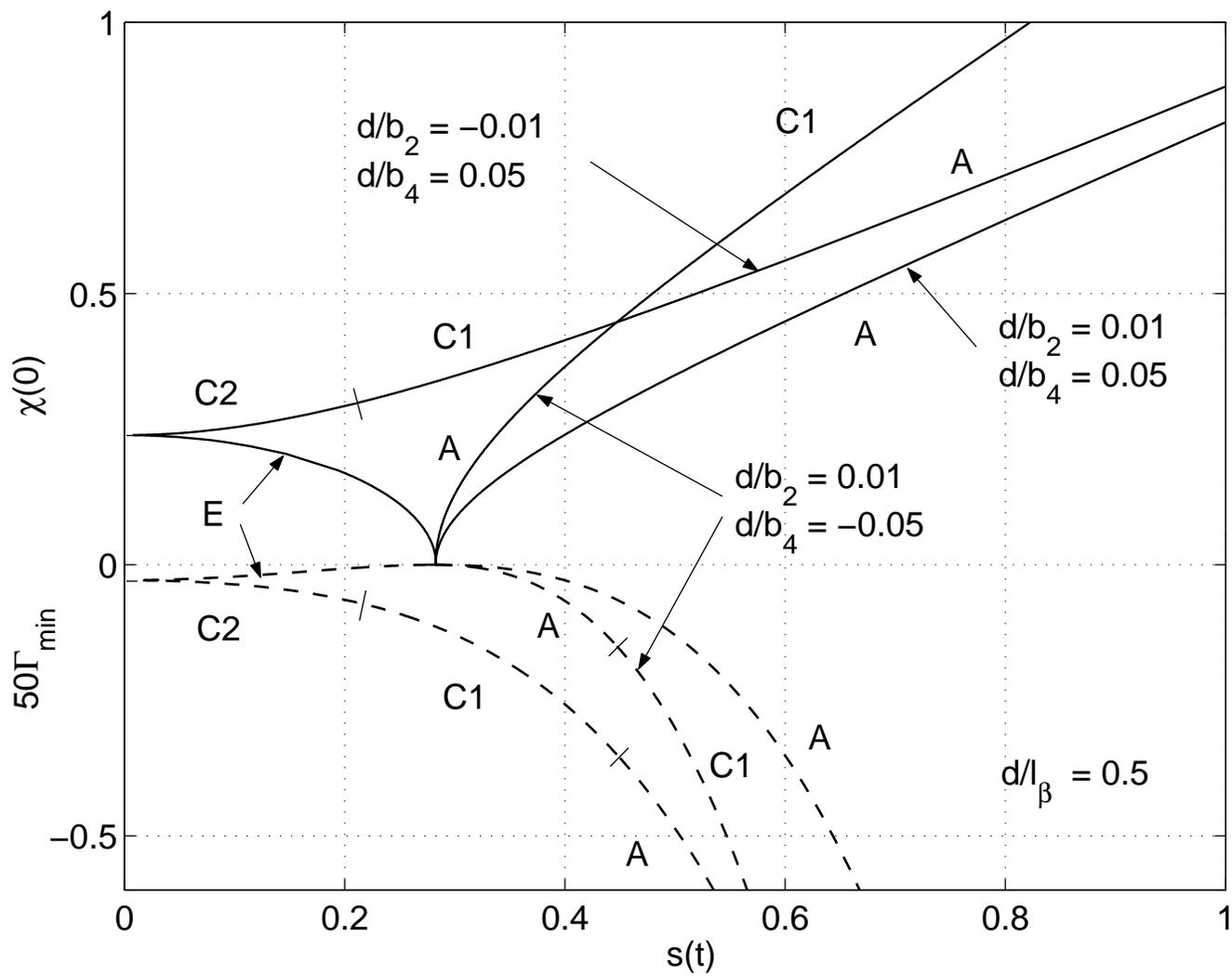
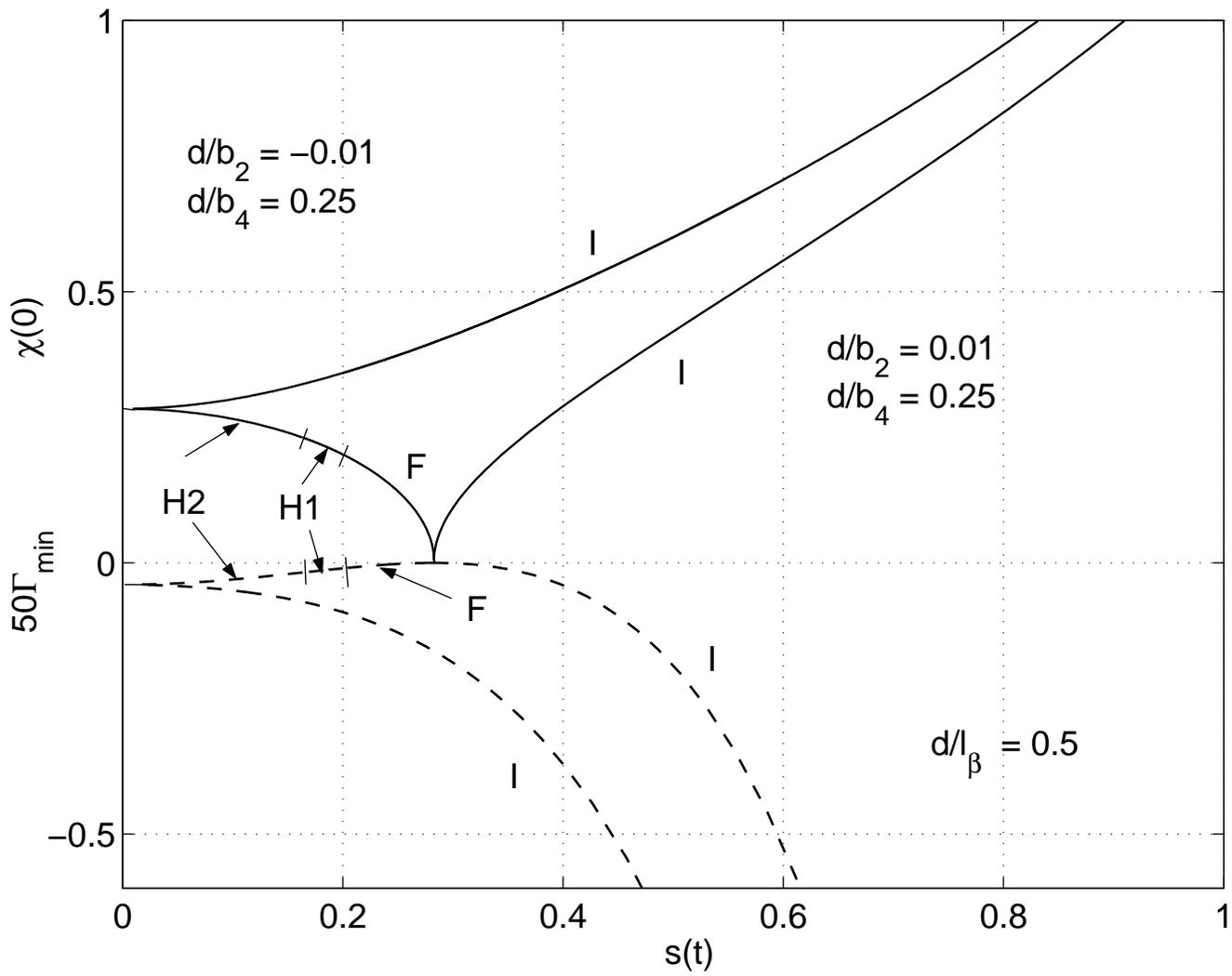
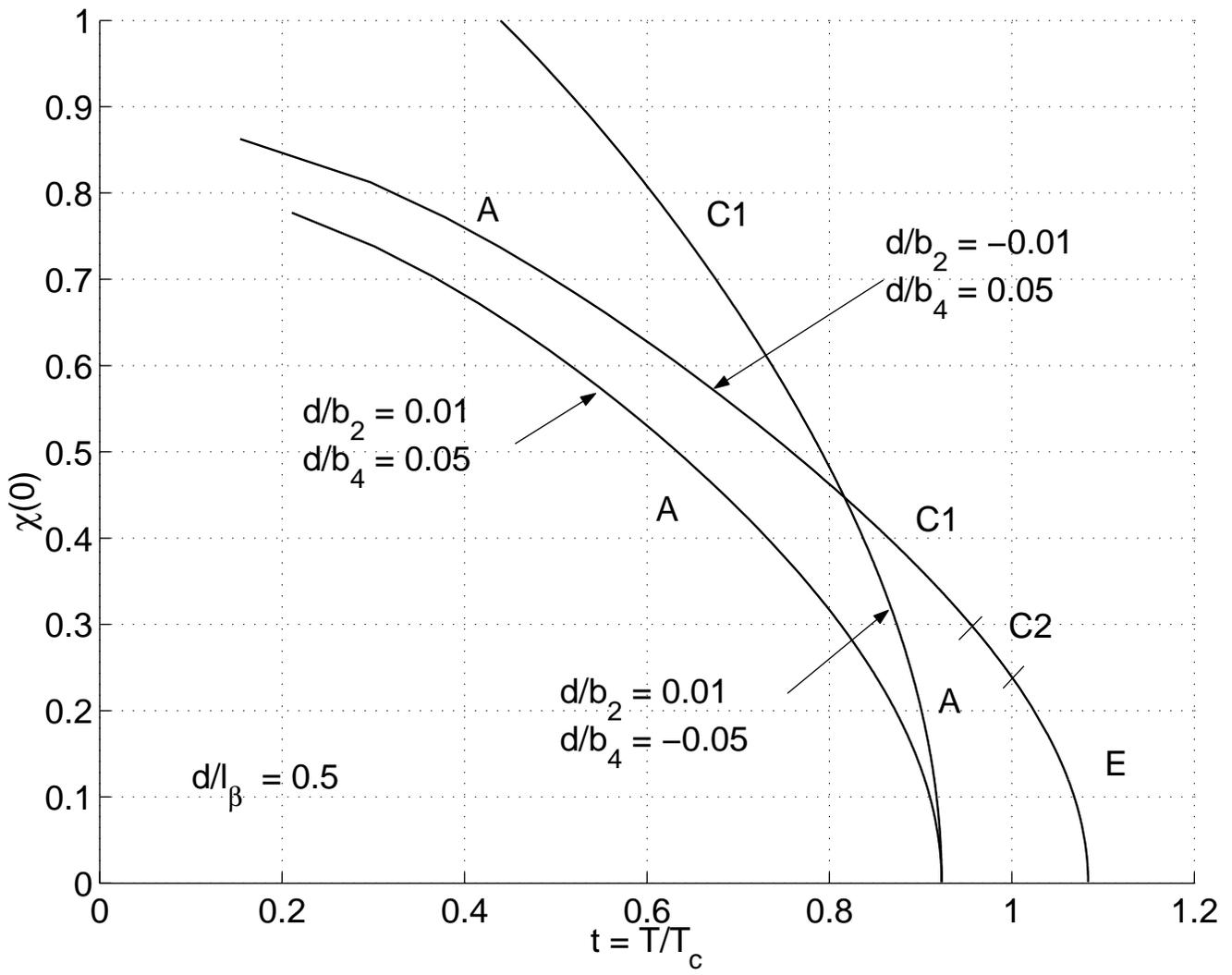
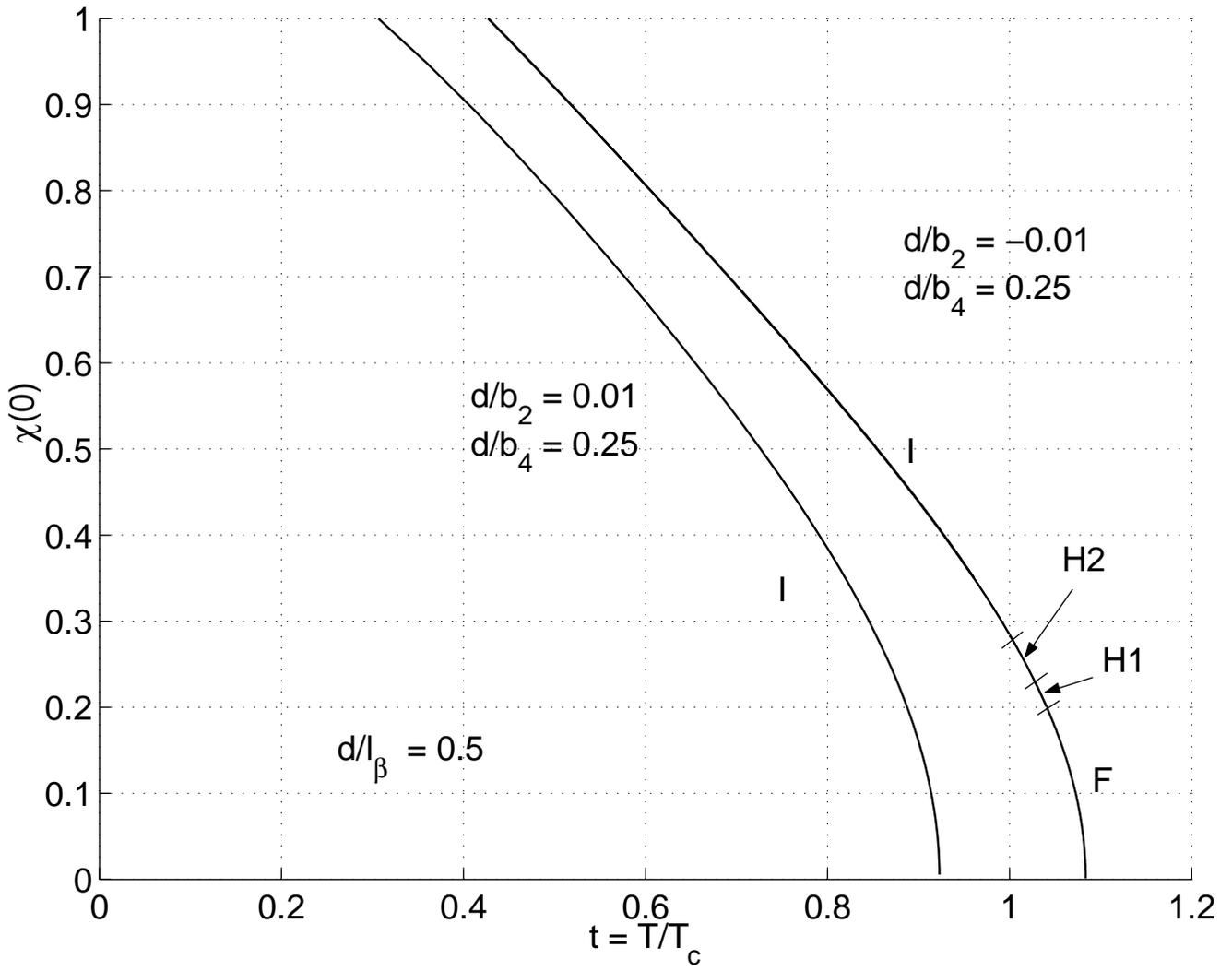
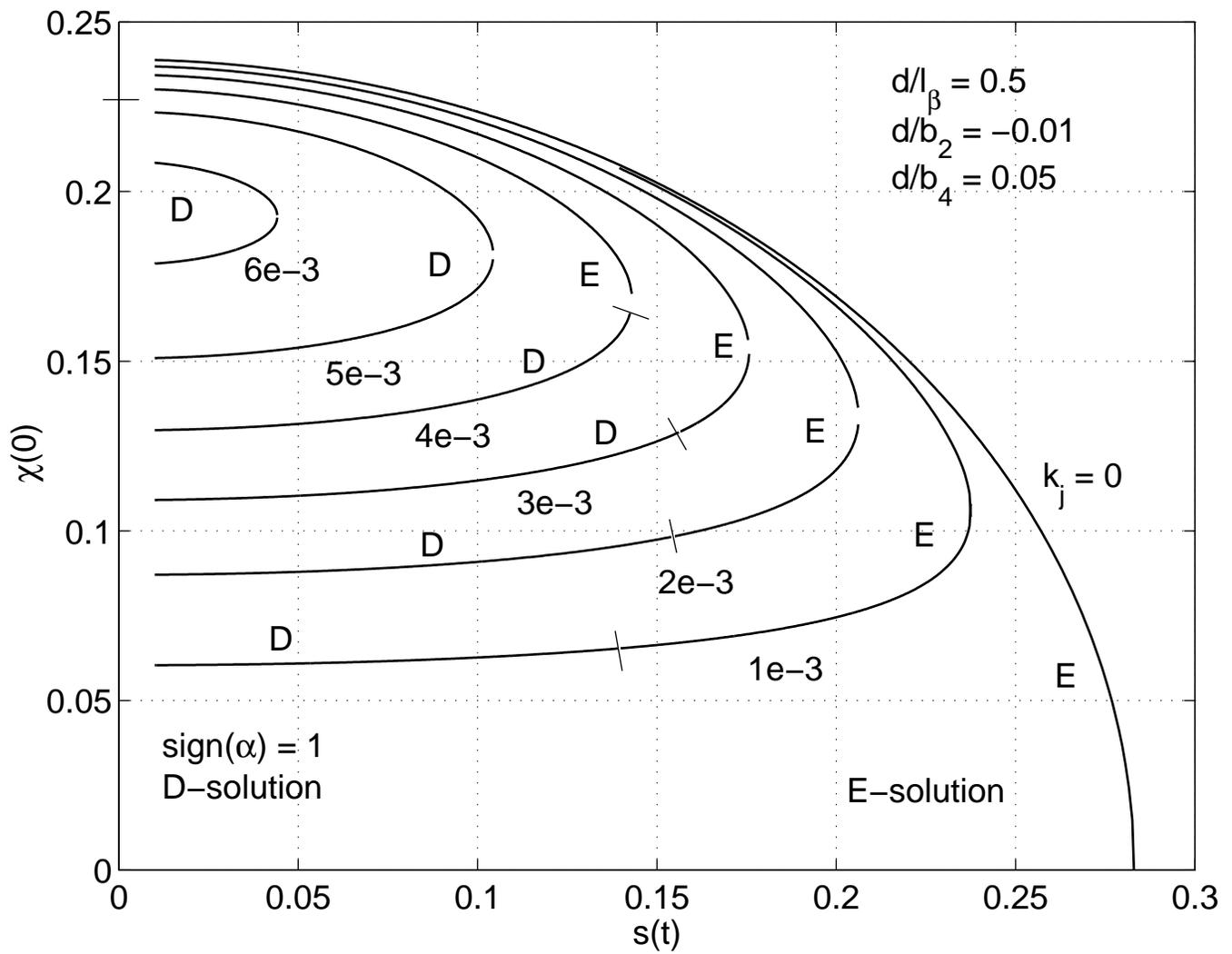
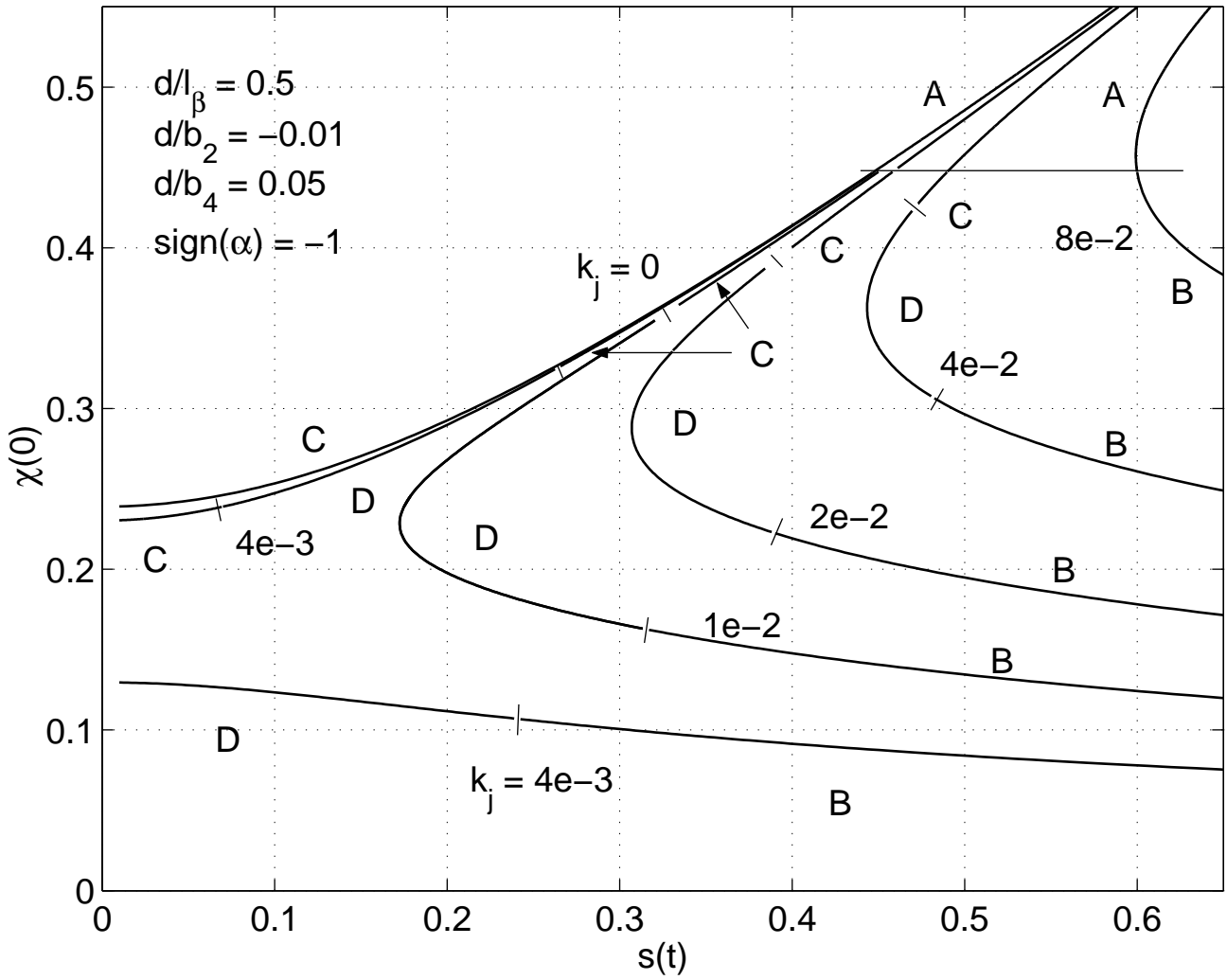
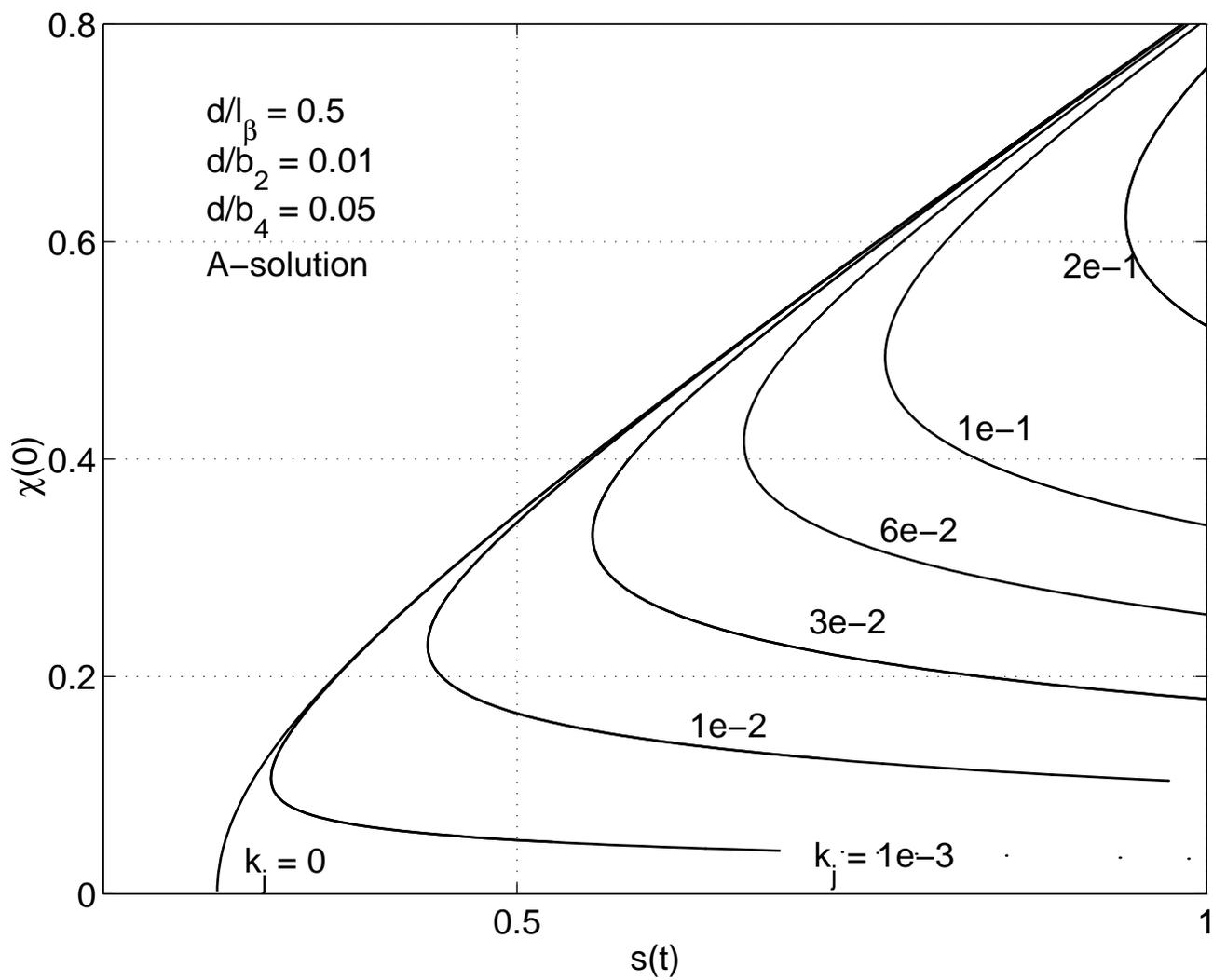
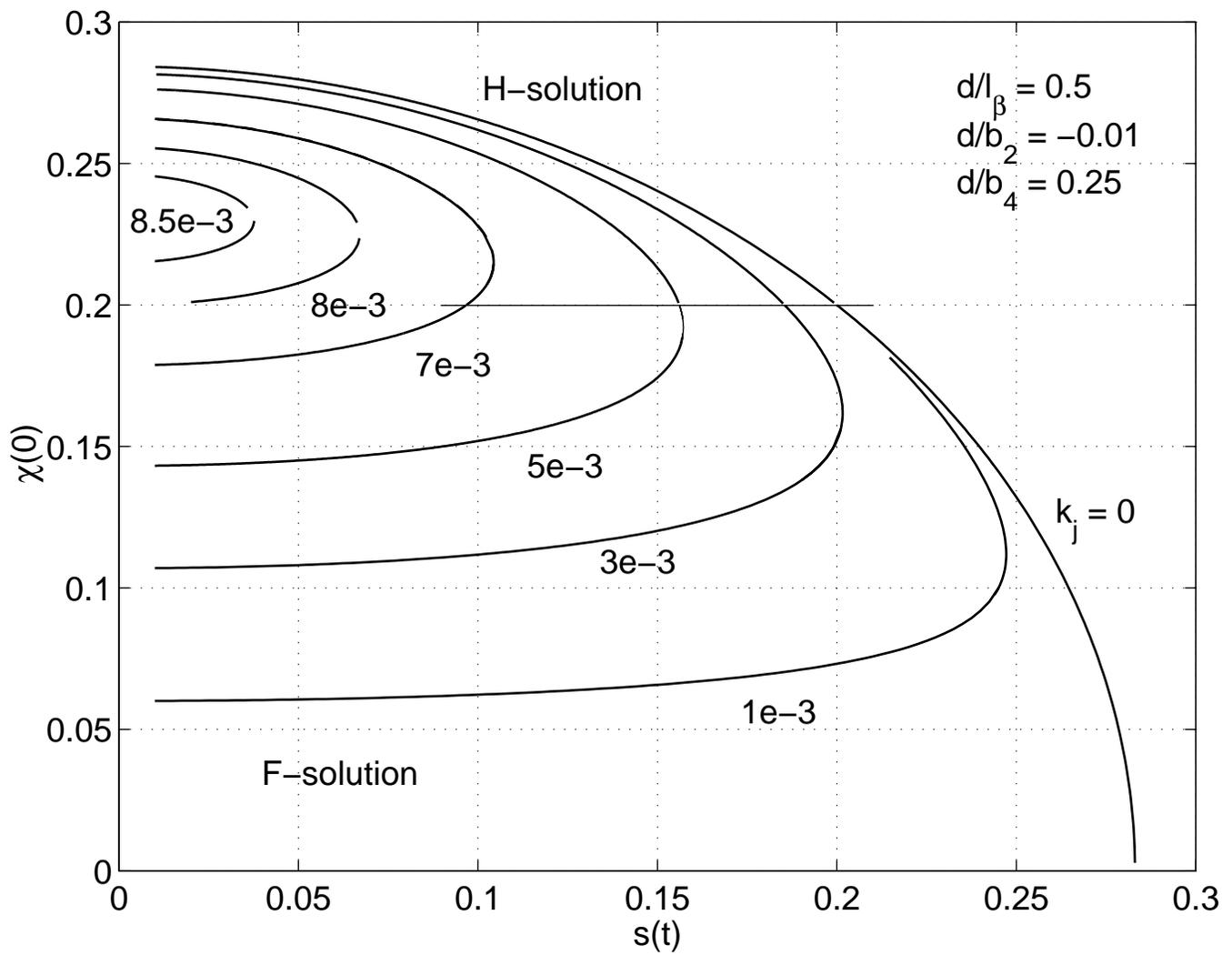
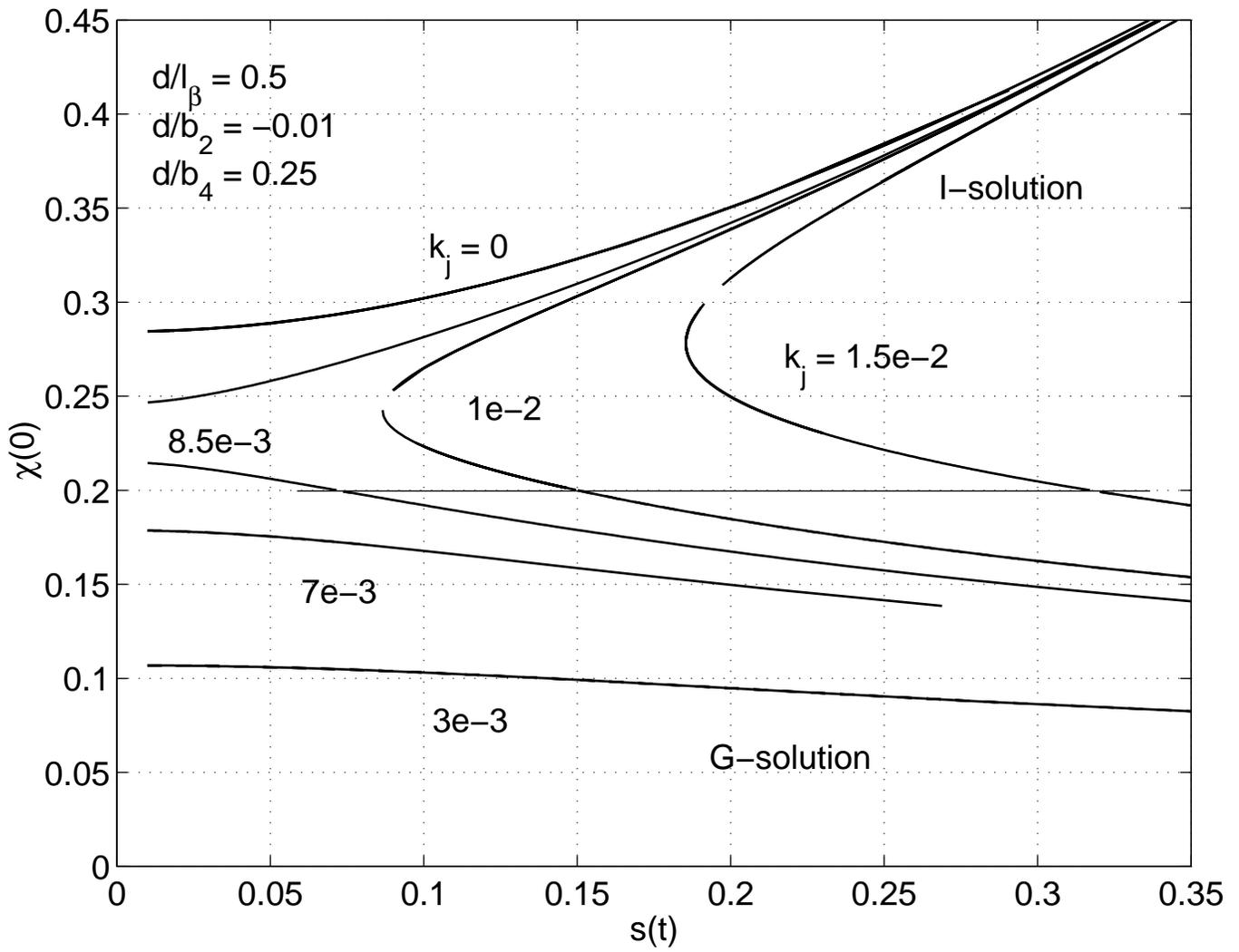
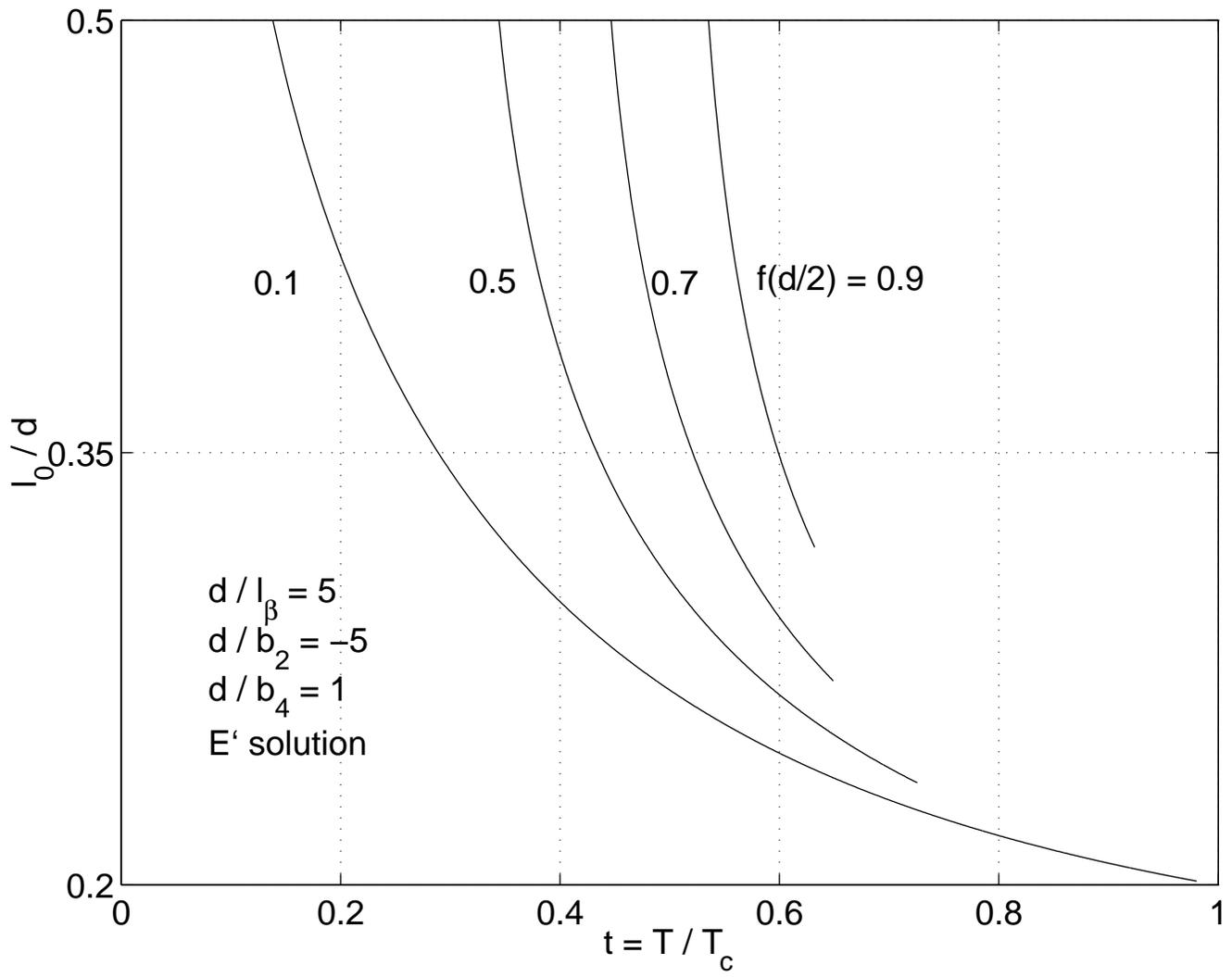
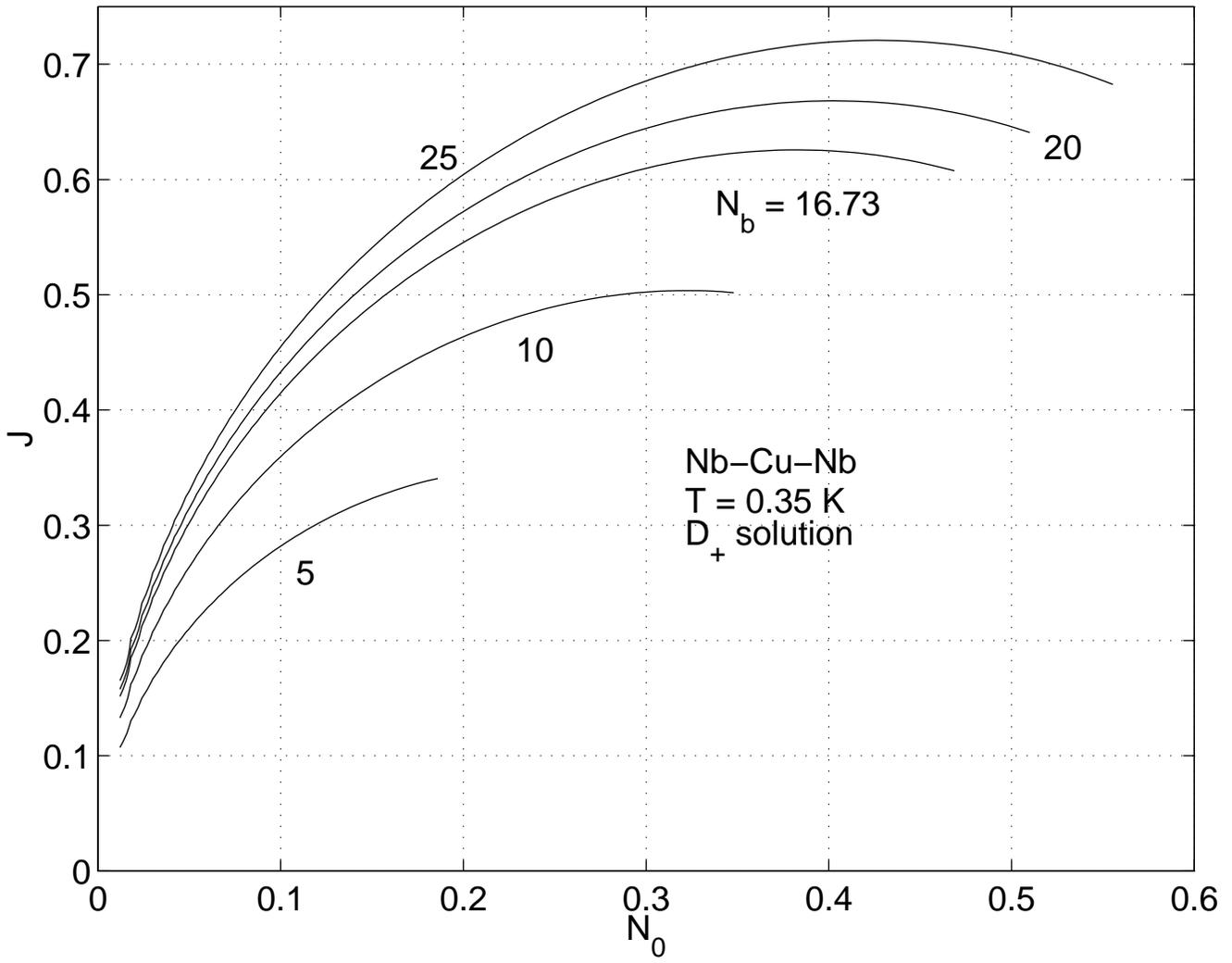
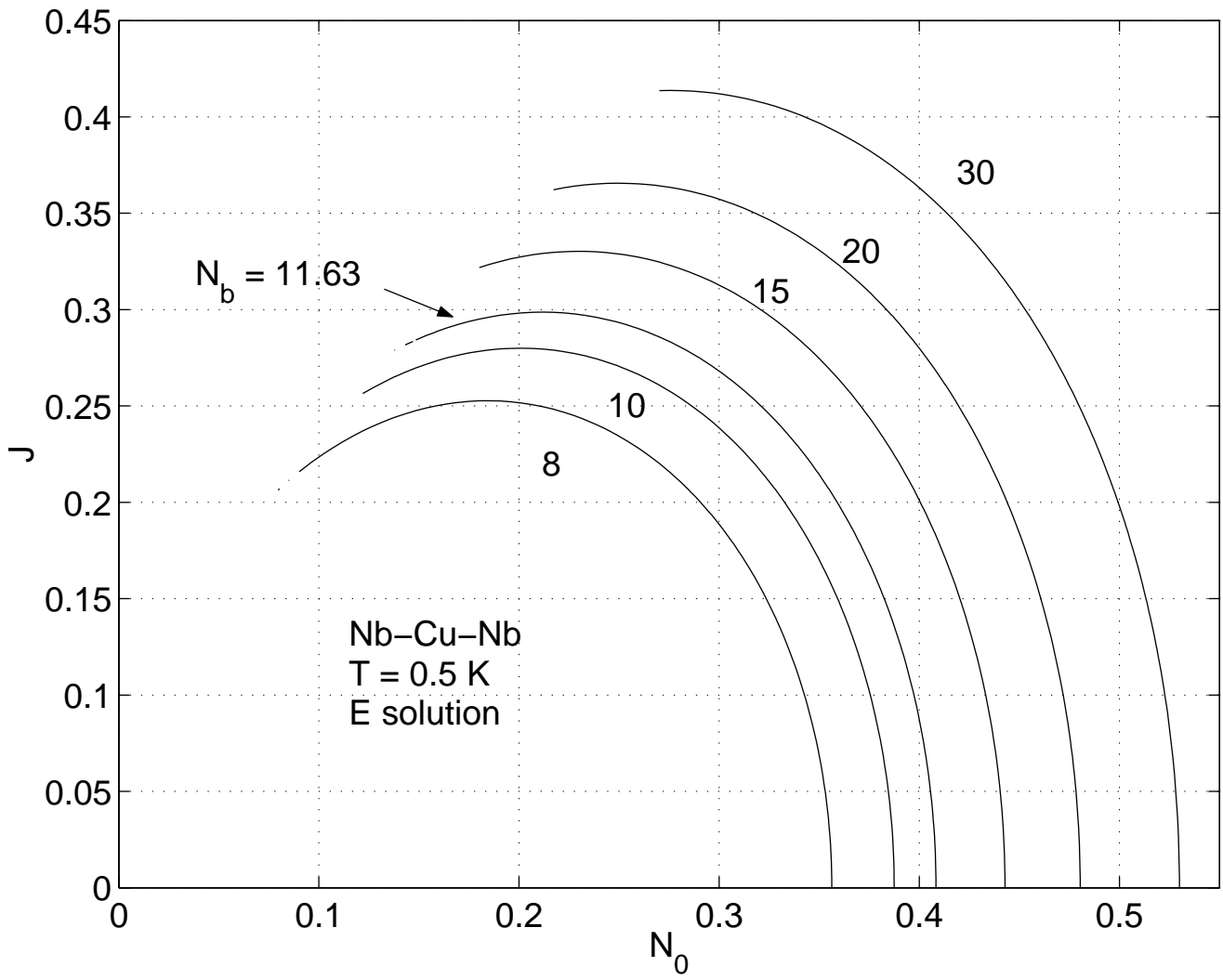
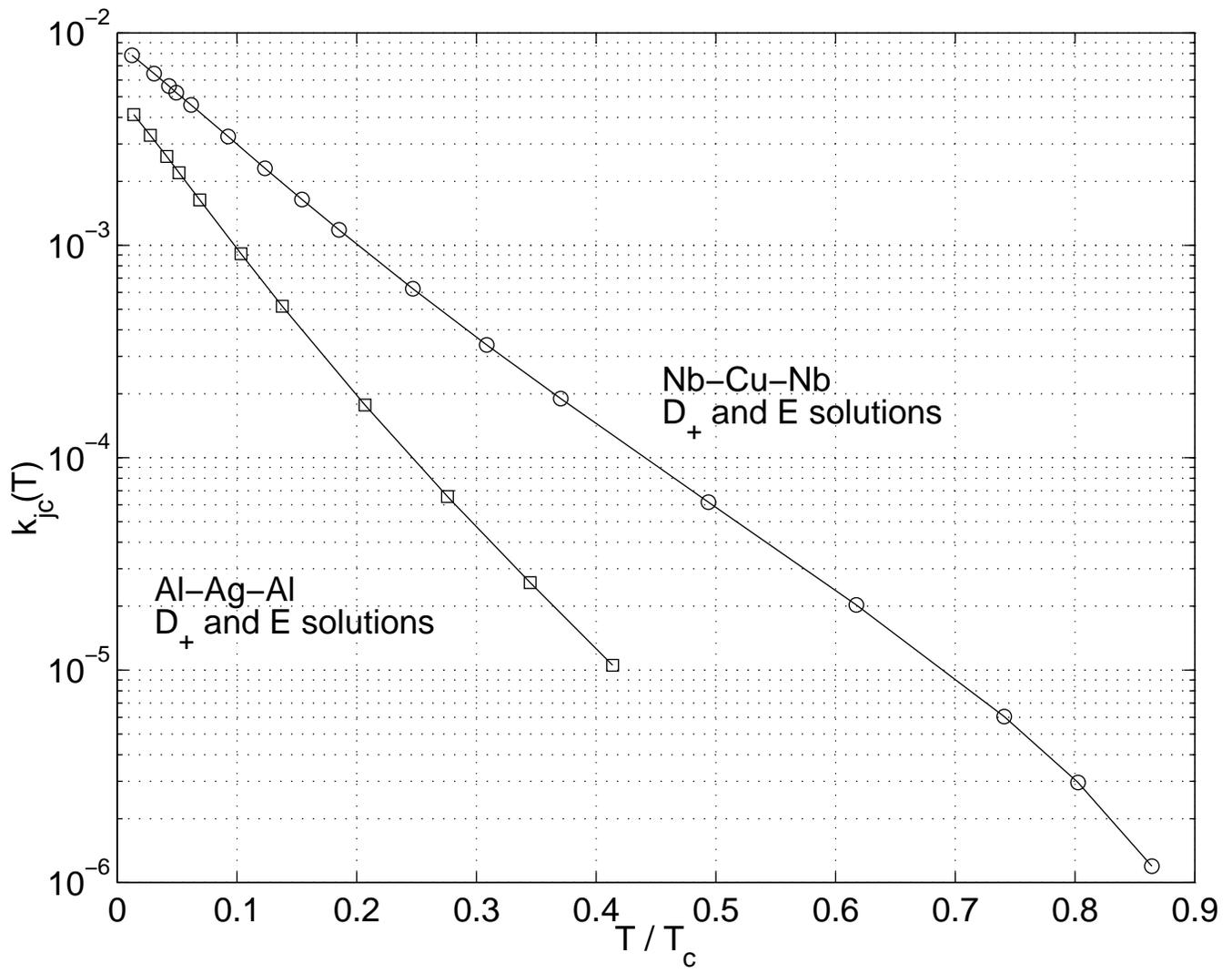
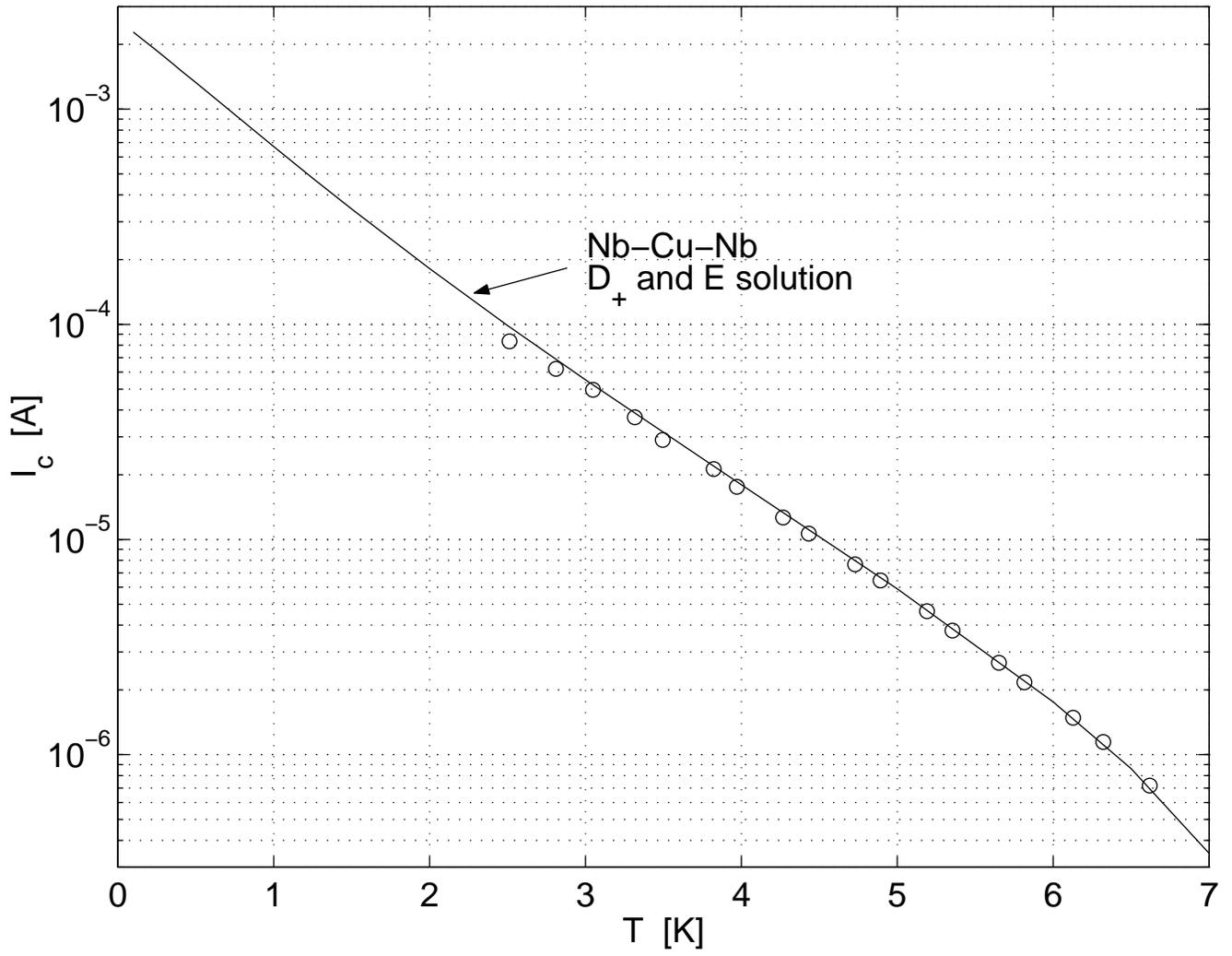
\end{document}